\documentclass[%
 reprint,
 amsmath,amssymb,
 aps,
]{revtex4-2}

\usepackage{graphicx}
\usepackage{dcolumn}
\usepackage{bm}
\usepackage{fancyvrb} 
\usepackage{breakurl}
\usepackage{multirow}
\usepackage{xcolor}
\VerbatimFootnotes    
\usepackage{enumitem}
\usepackage{xspace}

\usepackage[mathlines]{lineno}

\begin{document}

\preprint{APS/123-QED}

\newcommand{\vdag}{(v)^\dagger}
\newcommand\aastex{AAS\TeX}
\newcommand\latex{La\TeX}
\newcommand{\jw}[1]{\textcolor{red}{[{\bf JW}: #1]}}
\newcommand{\revision}[1]{\textcolor{blue}{{\bf #1}}}
\newcommand{\txs}{TXS~0506+056\xspace}
\newcommand{\nustar}{\textit{NuSTAR}\xspace}
\newcommand{\gb}{GB6\,J1542+6129\xspace} 
\newcommand{\cgcg}{CGCG\,420-015\xspace} 
\newcommand{\fermi}{\textit{Fermi}\xspace}
\newcommand{\event}{IceCube-170922A\xspace}
\hyphenation{Nu-STAR}
\newcommand{\flux}{\,erg\,s$^{-1}$\,cm$^{-2}$\xspace}
\newcommand{\ergs}{\,erg\,s$^{-1}$\xspace}

\newcommand*\aap{A\&A}
\let\astap=\aap
\newcommand*\aapr{A\&A~Rev.}
\newcommand*\aaps{A\&AS}
\newcommand*\actaa{Acta Astron.}
\newcommand*\aj{AJ}
\let\applopt\ao
\newcommand*\apjl{ApJ}
\let\apjlett\apjl
\newcommand*\apjs{ApJS}
\let\apjsupp\apjs
\newcommand*\aplett{Astrophys.~Lett.}
\newcommand*\apspr{Astrophys.~Space~Phys.~Res.}
\newcommand*\apss{Ap\&SS}
\newcommand*\araa{ARA\&A}
\newcommand*\azh{AZh}
\newcommand*\baas{BAAS}
\newcommand*\bac{Bull. astr. Inst. Czechosl.}
\newcommand*\bain{Bull.~Astron.~Inst.~Netherlands}
\newcommand*\caa{Chinese Astron. Astrophys.}
\newcommand*\cjaa{Chinese J. Astron. Astrophys.}
\newcommand*\fcp{Fund.~Cosmic~Phys.}
\newcommand*\gca{Geochim.~Cosmochim.~Acta}
\newcommand*\grl{Geophys.~Res.~Lett.}
\newcommand*\iaucirc{IAU~Circ.}
\newcommand*\icarus{Icarus}
\newcommand*\jcap{J. Cosmology Astropart. Phys.}
\newcommand*\jgr{J.~Geophys.~Res.}
\newcommand*\jqsrt{J.~Quant.~Spectr.~Rad.~Transf.}
\newcommand*\jrasc{JRASC}
\newcommand*\memras{MmRAS}
\newcommand*\memsai{Mem.~Soc.~Astron.~Italiana}
\newcommand*\mnras{MNRAS}
\newcommand*\na{New A}
\newcommand*\nar{New A Rev.}
\newcommand*\nphysa{Nucl.~Phys.~A}
\newcommand*\pasa{PASA}
\newcommand*\pasj{PASJ}
\newcommand*\pasp{PASP}
\newcommand*\physrep{Phys.~Rep.}
\newcommand*\physscr{Phys.~Scr}
\newcommand*\planss{Planet.~Space~Sci.}
\newcommand*\procspie{Proc.~SPIE}
\newcommand*\qjras{QJRAS}
\newcommand*\rmxaa{Rev. Mexicana Astron. Astrofis.}
\newcommand*\skytel{S\&T}
\newcommand*\solphys{Sol.~Phys.}
\newcommand*\sovast{Soviet~Ast.}
\newcommand*\ssr{Space~Sci.~Rev.}
\newcommand*\zap{ZAp}

\title{Possible correlation between unabsorbed hard X-rays and neutrinos in radio-loud and radio-quiet AGN}

\author{Emma Kun}
\affiliation{Astronomical Institute, Faculty for Physics \& Astronomy, Ruhr University Bochum, 44780 Bochum, Germany}
\affiliation{Theoretical Physics IV: Plasma-Astroparticle Physics, Faculty for Physics \& Astronomy, Ruhr University Bochum, 44780 Bochum, Germany}
\affiliation{Ruhr Astroparticle And Plasma Physics Center (RAPP Center), Ruhr-Universit\"at Bochum, 44780 Bochum, Germany}
\affiliation{Konkoly Observatory, HUN-REN Research Centre for Astronomy and Earth Sciences, Konkoly Thege Miklós \'ut 15-17, H-1121 Budapest, Hungary}
\affiliation{CSFK, MTA Centre of Excellence, Konkoly Thege Miklós \'ut 15-17, H-1121 Budapest, Hungary}
\email{ekun@astro.ruhr-uni-bochum.de}

\author{Imre Bartos}
\affiliation{Department of Physics, University of Florida, PO Box 118440, Gainesville, FL 32611-8440, USA}
\email{imrebartos@ufl.edu}

\author{Julia Becker Tjus}
\affiliation{Theoretical Physics IV: Plasma-Astroparticle Physics, Faculty for Physics \& Astronomy, Ruhr University Bochum, 44780 Bochum, Germany}
\affiliation{Ruhr Astroparticle And Plasma Physics Center (RAPP Center), Ruhr-Universit\"at Bochum, 44780 Bochum, Germany}
\affiliation{Department of Space, Earth and Environment, Chalmers University of Technology, SE-412 96 Gothenburg, Sweden}

\author{Peter L.\ Biermann}
\affiliation{MPI for Radioastronomy, 53121 Bonn, Germany}
\affiliation{Department of Physics \& Astronomy, University of Alabama, Tuscaloosa, AL 35487, USA}

\author{Anna Franckowiak}
\affiliation{Astronomical Institute, Faculty for Physics \& Astronomy, Ruhr University Bochum, 44780 Bochum, Germany}

\author{Francis Halzen}
\affiliation{Deptartment of Physics, University of Wisconsin, Madison, WI 53706, USA}

\author{Santiago del Palacio}
\affiliation{Department of Space, Earth and Environment, Chalmers University of Technology, SE-412 96 Gothenburg, Sweden}

\author{Jooyun Woo}
\affiliation{Columbia Astrophysics Laboratory, 550 West 120th Street, New York, NY 10027, USA}

\date{\today}

\begin{abstract}
The first high-energy neutrino source identified by IceCube was a blazar -- an active galactic nucleus driving a relativistic jet towards Earth. Jets driven by accreting black holes are commonly assumed to be needed for high-energy neutrino production. Recently, IceCube discovered neutrinos from Seyfert galaxies, which appears unrelated to jet activity. Here, we show that the observed luminosity ratios of neutrinos and hard X-rays from blazars TXS\,0506+056 and \gb are consistent with neutrino production in a $\gamma$-obscured region near a central supermassive black hole, with the X-ray flux corresponding to reprocessed $\gamma$-ray emission with flux comparable to that of neutrinos. Similar neutrino--hard X-ray flux ratios are found for four Seyfert galaxies, NGC\,1068, NGC\,4151, CGCG\,420-015 and NGC\,3079, raising the possibility of a common neutrino production mechanism that may not involve a strong jet. 
\end{abstract}

\maketitle


\section{Introduction}
\label{section:intro}

The IceCube Neutrino Observatory has already made a series of transformational discoveries. These include a quasi-diffuse flux of cosmic neutrinos of so-far unknown origin \cite{2013PhRvL.111b1103A}, as well as a growing number of individual astrophysical sites with associated neutrino emission. 

The first individual neutrino source identified at $>3\sigma$ significance was the blazar TXS 0506+056 \citep{ICTXS2018a}. Blazars represent a special class of active
galactic nuclei (AGN) that drive relativistic jets that point directly towards Earth. 
Relativistic jets can be sources of high-energy neutrinos if their accelerated protons undergo $p\gamma$ or $pp$ interactions \citep[see e.g., ][]{msb1992,bbr2005,physirep2008,radiogalaxies2014}, however, for the latter external material \citep[][]{2010A&A...522A..97A} or structured jets are needed \citep[][]{2015MNRAS.451.1502T}. The second neutrino source identified with high confidence was the Seyfert galaxy NGC\,1068 \citep{ngc1068_2022}. Seyfert galaxies are radio-quiet AGNs with much weaker jets compared to  blazars. Their hard X-ray emission likely originates from the hot plasma called corona surrounding the accretion disk of the central black hole in AGN. 

Subsequently, additional astrophysical sources have been associated with neutrino emission. These included two blazars (PKS\,1502+106, \citep{2019ATel12967....1T}; PKS\,1424$-$41, \citep{2016NatPh..12..807K}) and three Seyfert galaxies \citep[NGC\,4151, NGC\,3079 \cgcg;][]{2023arXiv230715349G,Neronov2023,Abassi2024}. 
Additional neutrino source candidates include the blazars PKS~1424+240 ($3.7\sigma$) and \gb ($2.2\sigma$) \citep{ngc1068_2022}. Similarly to TXS 0506+056, these two blazars also possibly belong to a class of masquerading BL Lac objects \citep{Padovani2022}.

High-energy neutrinos are produced through the interactions of relativistic protons accelerated in the sources. These interactions also produce $\gamma$-rays with comparable flux and energy spectra to that of neutrinos. However, if such interactions take place near the central black hole, where plenty of infrared-optical photons from the accretion disk and X-rays from the hot corona are present, active pair-production will convert, or reprocess, the $\gamma$-rays into $\lesssim 1$\,MeV photons (hard X-rays) \citep{2020PhRvL.125a1101M,2020ApJ...891L..33I,2022ApJ...939...43E,Neronov2023}. The apparent underproduction of $\gamma$-rays compared to neutrinos in Seyfert galaxies is naturally explained in such environments \citep[e.g.,][]{2021ApJ...911L..18K,Kun2023}. Even blazars, generally bright $\gamma$-ray sources, tend to show a temporary $\gamma$-opaqueness during neutrino emission. A hint of $\gamma$-absorption is also observed in the diffuse neutrino flux \citep{PhysRevLett.116.071101,2020PhRvL.125a1101M}.

Consequently, hard X-ray emission produced by this reprocessing should have comparable flux to that of high-energy neutrinos. The linear scaling between the \textit{unabsorbed} hard X-ray and neutrino fluxes for Seyfert galaxies has already been suggested by \cite{2016PhRvD..94j3006M}. Based on this linear scaling and observations in the 2--10\,keV band, \citet{2020PhRvL.125a1101M} listed the brightest neutrino source candidate AGNs including NGC 1068. The AGNs NGC\,1068 \citep[$z=0.003810$,][]{2004MNRAS.350.1195M}, NGC\,3079 \citep[$z=0.00399$,][]{2022ApJS..261....6K} and NGC\,4151\citep[$z=0.003152$,][]{2022ApJS..261....6K} were indeed suggested to have comparable \textit{unabsorbed} hard X-ray (15--195\,keV range) and all-flavor neutrino fluxes \citep{Neronov2023}.

\section{Hard X-ray fluxes}

Due to significant reprocessing by multiple Compton scatterings and photoelectric absorption in Seyfert galaxies, complicated obscurer models (e.g., \texttt{MYTorus} \citep{mytorus_tbabs,mytorus}) are often adopted to account for such reprocessing of X-rays and calculate the \textit{unabsorbed} flux. \citet{Neronov2023}, however, used a simple exponential absorption correction and a power-law model to convert the \textit{observed} hard X-ray flux.

We estimated the \textit{unabsorbed} hard X-ray flux of the four IceCube-detected Seyfert galaxies, namely NGC 1068, NGC 4151, NGC 3079 and \cgcg to deduce a physically appropriate relation between their hard X-ray and neutrino flux. We extended this comparison to the all-time average of the blazar \txs \citep[$z=0.3365$,][]{2018ApJ...854L..32P}, the only blazar with neutrino flux identified with $>3\sigma$ confidence. 

We additionally carried out an X-ray observation of the blazar \gb using the NuSTAR X-ray satellite. \gb is another blazar associated with IceCube neutrinos. We corroborate that the X-ray absorption in the Galactic interstellar medium is significantly smaller than the intrinsic absorption for these AGN, the Galactic hydrogen column density being in the order of $\sim 10^{20}~\mathrm{cm}^{-2}$ \citep{1990ARA&A..28..215D,2005A&A...440..775K,2016A&A...594A.116H}.


\subsection{Seyferts: NGC~1068, NGC~3079, NGC~4151 and CGCG 420-015}

Regarding the hard X-ray flux of NGC\,1068, we extrapolated the 10--40 keV Nuclear Spectroscopic Telescope Array \citep[\nustar,][]{2013ApJ...770..103H} unabsorbed luminosity $L_\mathrm{10-40,intr} \approx 1.5\times 10^{43}~\mathrm{erg}~\mathrm{s}^{-1}$ from \citet{2015ApJ...812..116B} to the 15--55 keV range, with the spectral index $\Gamma= 2.10\pm0.07$. The resulted luminosity of the main continuum is $L_\mathrm{15-55,intr} \approx 1.36\pm0.15\times 10^{43}~\mathrm{erg}~\mathrm{s}^{-1}$, where we assumed a 10\% error on the 10--40 keV luminosity.

For NGC~3079, assuming an edge-on configuration, \citet{Marchesi2018} favors a value of $2.47\pm0.24 \times 10^{24}~\mathrm{cm}^{-2}$ for $N_H$ (along the line-of-sight), which is in agreement with the value $N_H=3.2^{+0.54}_{-0.43}\times 10^{24} \mathrm{cm}^{-2}$ suggested by \citet{2019A&A...621A..28G}, also relying on \nustar observations. We adopt the luminosity of the main continuum of NGC\,3079 of $L_\mathrm{15-55,intr}=2.63^{+0.61}_{-0.59} \times 10^{42}~\mathrm{erg}~\mathrm{s}^{-1}$,  
from \citet{Marchesi2018}.

In the case of NGC~4151, the favored in-source column density is $N_H\sim 10^{22}$--$10^{23}~\mathrm{cm}^{-2}$ \citep[e.g.][]{1992MNRAS.259..369P,2019ApJ...884...26Z}, or even smaller.  \citet{2023MNRAS.523.4468G} gives the parameters for the Comptonized primary continuum of NGC~4151; we retrieved the 15--55 keV flux by loading in \texttt{XSPEC} the \texttt{nthcomp} model with their fitted parameters, obtaining an unabsorbed flux of $F_\mathrm{15-55} \approx 3.1\times 10^{-10}$\flux. 

The unabsorbed hard X-ray luminosity of \cgcg \citep[z=0.0296,][]{2022ApJS..261....2K} was estimated by \citet{Marchesi2018} as $\log (L_\mathrm{15-55,intr})=43.65^{+0.09}_{-0.12}~\mathrm{erg}~\mathrm{s}^{-1}$ with a line-of-sight column density of $N_H=7.15^{+0.85}_{-0.97}~\times 10^{23}~\mathrm{cm}^{-2}$.

\subsection{Blazar TXS 0506+056\label{subsec:txs_nustar}}

As a broadband (3--79 keV) focused (FWHM $14"$) hard X-ray space telescope, \nustar has been providing unique opportunities to study extreme phenomena of AGNs in the X-ray band including \txs. We analyzed all available \nustar data of \txs, consisting of 18 observations between 2017 and 2021 with a total exposure of 371~ks. We processed the data using NuSTAR Data Analysis Software (NuSTARDAS v2.1.2) and CALDB 20230718. We generated cleaned data products using \texttt{nupipeline} task with \texttt{saamode=strict} and \texttt{tentacle=yes} flags. Source (background) spectra were extracted from a circular region with radius $30"$ (annular region with radius $1'.5-2'.5$) centered at \txs. The source is bright over the background in 3 to $\sim 40$\,keV for all observations. We modeled the spectra from each observation separately with an absorbed red-shifted power law (\texttt{tbabs*zpow}) in \texttt{XSPEC} \citep{1996ASPC..101...17A} using the Galactic hydrogen column density  ($N_{\textrm{H}}=1.55\times10^{21}$ cm$^{-2}$), the abundance table from \citet{wilm} (\texttt{abund wilm}), and redshift $z=0.3365$ \citep{2018ApJ...854L..32P}. The power law index varies between 1.5 -- 1.9 among observations. The source is variable in the hard X-ray roughly within a factor of $\sim 2$ as reported in \citet{variability}. The list of observations and 15--55 keV luminosity for each observation is shown in Table \ref{tab:nustar}. To provide observational window for the X-ray luminosity comparable to the integrated neutrino luminosity, we calculated the luminosity of \txs{} by averaging the luminosity measurement from all the available observations. The average 15--55 keV luminosity of \txs is $(9.0\pm2.4)\times10^{44}$\ergs, where we adopted the $1\sigma$ quartile of the flux distribution as the error-bar to account for variability.

\begin{table}
\caption{\nustar observations used to calculate the hard X-ray vs. high-energy neutrino flux relation. \label{tab:nustar}}
\begin{tabular}{cccc}
\hline
\hline
{ObsID} & {Date} & {Exposure} & {$L_{15-55}$} \\
{} & {} & {(ks)} & ($10^{44}$\ergs)\\
\hline
90301618002 & 2017-09-29 & 21.6 & $6.5\pm0.8$ \\
90301618004 & 2017-10-19 & 19.7 & $6.6\pm0.9$ \\
90401610002 & 2018-04-03 & 2.2 & $10.6\pm1.1$ \\
90402637002 & 2018-10-16 & 26.5 & $10.7\pm0.9$ \\
90402637004 & 2018-11-15 & 22.9 & $7.0\pm0.9$ \\
90402637006 & 2018-12-08 & 20.8 & $8.3\pm0.9$ \\
90402637008 & 2019-01-07 & 1.9 & $10.9\pm1.0$ \\
60502053002 & 2019-07-30 & 17.4 & $13.3\pm1.2$ \\
60502053004 & 2019-09-29 & 25.9 & $10.1\pm0.9$ \\
60502053006 & 2019-11-29 & 17.7 & $10.8\pm1.2$ \\
60502053008 & 2020-01-26 & 15.1 & $9.5\pm1.2$ \\
60502053010 & 2020-03-25 & 20.8 & $12.3\pm1.1$ \\
60602004002 & 2020-09-26 & 18.3 & $9.7\pm1.1$ \\
60602004004 & 2020-10-25 & 22.0 & $12.3\pm1.2$\\
60602004006 & 2020-11-17 & 20.4 & $7.9\pm0.9$ \\
60602004008 & 2020-12-10 & 20.1 & $6.7\pm0.9$ \\
60602004010 & 2021-01-16 & 17.5 & $4.0\pm0.8$\\
60602004012 & 2021-02-12 & 22.8 & $5.1\pm0.7$\\
\hline
\multicolumn{2}{c}{Total exposure} & 370.7 & \\
\hline
\multicolumn{3}{c}{Average luminosity in 15--55 keV} & $9.0\pm0.2$ \\
\hline
\hline
\end{tabular}
\end{table}

\subsection{Blazar \gb}\label{subsec:gb6_nustar}

The blazar \gb ($z=0.507$, \citep{Marcha2013}) was observed first time by \nustar Program 10049 (PI del Palacio) with a 36~ks exposure. We generated cleaned data products using \texttt{nupipeline} task with \texttt{saacalc=2}, \texttt{saamode=optimized} and \texttt{tentacle=yes} flags. The source spectrum was extracted from a circular region with radius $30"$ centered on \gb and the background was chosen from a larger ellipsoidal region in a source-free region within the same chip. \gb is detected significantly above the background up to $\sim$20~keV. As done with \txs, we modeled the spectrum with an absorbed red-shifted power law (\texttt{tbabs*zpow}) in \texttt{XSPEC}, although the absorption is negligible due to the low column density towards this source ($N_{\textrm{H}}=0.13\times10^{21}$\,cm$^{-2}$). The power-law photon index is $\Gamma = 1.55 \pm 0.15$, which is significantly harder than the spectral index inferred from \textit{Swift}-XRT observations ($\Gamma \sim 2.5$).
The model yields a flux in the 15--55~keV energy range of $F_{15-55}=(6.0\pm1.2)\times 10^{-13}$\flux. We note that the light curve obtained with \textit{Swift}-XRT \citep{Evans2007} for this sources does not exhibit strong flaring variability, suggesting that the X-ray emission from \gb is relatively steady within a factor $\sim$2, and therefore the reported flux should be representative of the source flux in the 10-yr IceCube window.

\section{High-energy neutrino fluxes}

\begin{figure*}
    \centering    
    \includegraphics[width=0.48\textwidth]{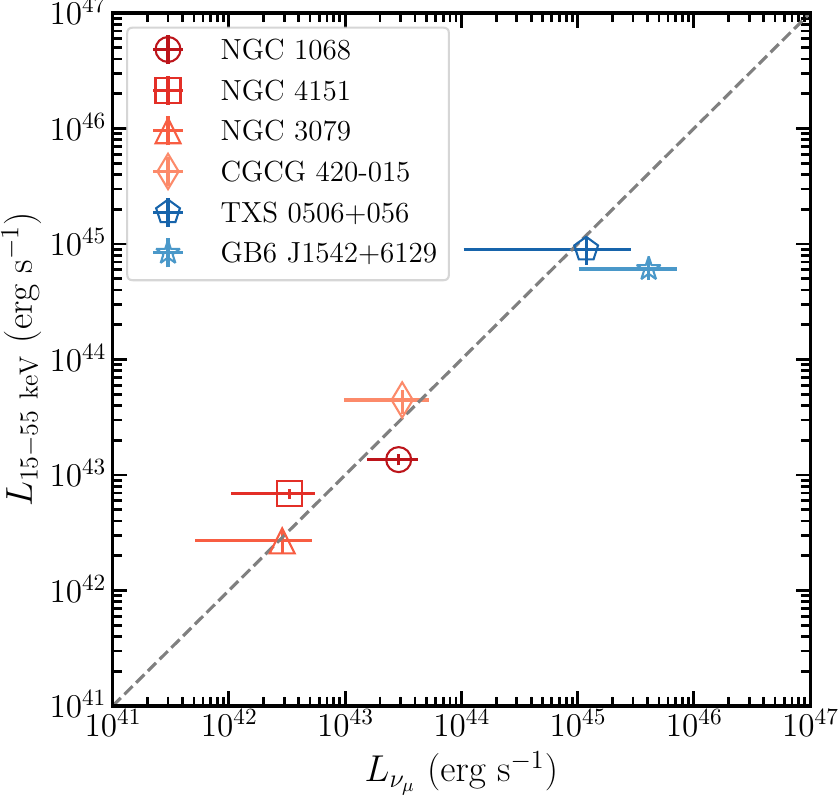} \hfill
    \includegraphics[width=0.48\textwidth]{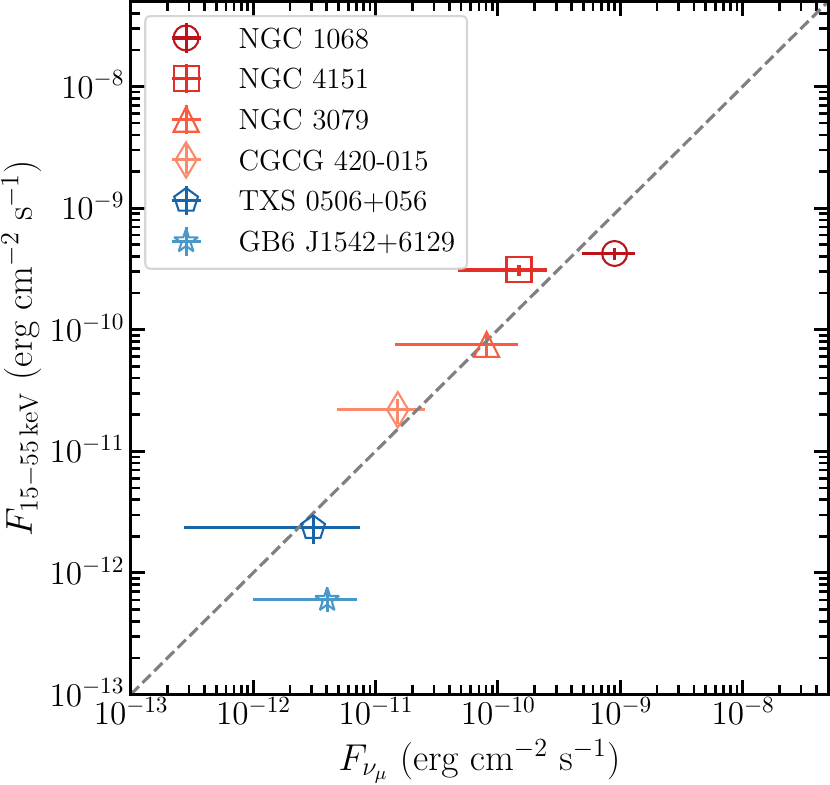} \\
    \caption{Correlation plot between the unabsorbed hard X-rays and neutrinos in radio-loud and radio-quiet AGN in our sample. Left: Luminosity of the main (unabsorbed) hard X-ray continuum shown as a function of the high-energy neutrino flux for four Seyferts, NGC~1068, NGC~4151, NGC~3079, CGCG~420-015 in red, and the blazars \txs and \gb in blue. The dashed line shows $L_{X}=L_{\nu}$. Right: the same as on the left plot, but in fluxes.}
    \label{fig:neronov-extended}
\end{figure*}

For three Seyfert galaxies NGC~1068, NGC~4151, NGC~3079, we assumed the high-energy neutrino fluxes reported by \citet{Neronov2023}. All neutrino spectra are assumed to be a power-law in the form of $\phi_\nu(E)=\phi_{0} (E_\nu/E_0)^{-\gamma}$,
where $\phi_{0}$ is the normalization factor and $\gamma$ is the spectral index ($E_0=1~\mathrm{TeV}$). We note all analyses use similar amount of neutrino data.

In the case of \cgcg, we estimated the $0.3$--$100$~TeV integrated neutrino flux from the best-fit normalization factor ($\phi_{0}\approx 1.2\times 10^{-11}~\mathrm{TeV}^{-1}~\mathrm{cm}^{-2}~\mathrm{s}^{-1}$) and spectral index ($\hat{\gamma}\approx2.8$) based on the likelihood scan from \citet{Abassi2024}. Then the integrated flux emerges as $F_{\nu_\mu + \bar{\nu}_\mu}=(3.8\pm2.5)\times 10^{-12}~\mathrm{TeV}~\mathrm{cm}^{-2}~\mathrm{s}^{-1}$, which comes from
\begin{equation}
        F_{\nu_\mu + \bar{\nu}_\mu} =\int^{E_2}_{E_1} E_\nu~ \phi_{\nu_\mu + \bar{\nu}_\mu} (E_\nu)~dE,
\end{equation}
with $E_1=0.3$\,TeV and $E_2=100$\,TeV.

We estimated the 0.3--100\,TeV integrated neutrino flux of \txs from \citet{ngc1068_2022}. 
We derived $\phi_{\nu_\mu + \bar{\nu}_\mu}$ such that  $\phi_{0}\approx 3.57\times10^{-13}~\mathrm{TeV}^{-1}~\mathrm{cm}^{-2}~\mathrm{s}^{-1}$ and $\hat{\gamma} \approx 2.04$. We estimated the relative error of the resulted flux considering the 68\% confidence interval at the 100\,TeV energy. Then the integrated neutrino flux between 0.3\,TeV and 100\,TeV results as $F_{\nu_\mu + \bar{\nu}_\mu} =1.93\times 10^{-12}~\mathrm{TeV}~\mathrm{cm}^{-2}~\mathrm{s}^{-1}$, with lower and upper limits of $1.70\times 10^{-13}~\mathrm{TeV}~\mathrm{cm}^{-2}~\mathrm{s}^{-1}$ and $4.70\times 10^{-12}~\mathrm{TeV}~\mathrm{cm}^{-2}~\mathrm{s}^{-1}$, respectively. 

In the case of \gb, the best-fit number of signal events ($\hat{n}_s=16.0$) and astrophysical spectral index ($\hat{\gamma}=3.0$) are published \citep{ngc1068_2022}. The number of signal events $n_{s}$ detected by IceCube from a source in direction of declination $\delta$ is given as:
\begin{equation}
n_{s}=\tau \int_0^\infty A_\mathrm{eff}(E_\nu,\delta) \times \phi_{\nu_\mu+\bar{\nu}_\mu}(E_\nu) dE
\end{equation}
where $\tau$ is the live-time, $A_\mathrm{eff}(E_\nu,\delta)$ is the energy and direction dependent effective area of the detector ($\delta_{\gb}=61.5^\circ$). The published tabulated effective areas of the IceCube Neutrino Detector range from $E_\mathrm{min}=100$ GeV to $E_\mathrm{max}=10$ PeV, in 40 log-scale bins \citep[seasons: IC40, IC59, IC79, IC86-I, IC86-II in][]{Abassi2019}. Then the number of signal events summing over in all seasons is
\begin{equation}
n_s= \phi_0 \sum_{n=1}^{5} \tau_n \left( \sum_{\mathrm{i=1}}^{40} A_\mathrm{eff,i} \int_{E_\mathrm{l,i}}^{E_\mathrm{u,i}} \left( \frac{E_\nu}{E_0} \right)^{-\gamma} dE\right)_n
\end{equation}
where $i$ the energy bin number running from $1$ to $40$, $\tau_1$--$\tau_5$ are the live-time of the seasons, $E_\mathrm{l,i}$($E_\mathrm{u,i})$ is the lower(upper) energy of the $i$th energy bin and $n$ the season number running from $1$ to $5$. Reorganizing the equation we get $\phi_0=1.5\times 10^{-12}~\mathrm{TeV}^{-1} ~\mathrm{cm}^{-2}~\mathrm{s}^{-1}$ (this is about the 90\% upper limit given by IceCube) and the 0.3--100 TeV integrated neutrino flux of \gb arises as $F_{\nu_\mu + \bar{\nu}_\mu}=5.0\pm3.8 \times 10^{-12}~\mathrm{TeV}~\mathrm{cm}^{-2}~\mathrm{s}^{-1}$. This flux is about 2.5 times larger compared to \txs in the 0.3--100 TeV range, which can be explained with the larger $\hat{n}_s$ and softer $\hat{\gamma}$ of \gb. We note that to be compatible with \citet{Neronov2023}, who gave the $F_{\nu_\mu}$, we did take account only for the muon component by halving the above fluxes.

\section{$\gamma$-obscured neutrino sources}

We show our results in the left side of Fig.~\ref{fig:neronov-extended}, where one can see a possible correlation between the unabsorbed hard X-ray and neutrino luminosities of the four Seyfert galaxies and the two blazar in our AGN sample. 
The same relation can be seen for between the hard X-ray and neutrino fluxes (right side of Fig.~\ref{fig:neronov-extended}, $H_0=69.6~\mathrm{km}~\mathrm{s}^{-1}~\mathrm{Mpc}^{-1}$, $\Omega_\mathrm{m,0}=0.286$, $\Omega_\mathrm{\lambda,0}=0.714$, $T_\mathrm{cmb,0}=2.72548~\mathrm{K}$ in a $\Lambda$CDM cosmology), which might be due to the still relatively small sample size. The Pearson correlation coefficient between the logarithm of luminosities emerges as $R=0.97$ hinting at strong linear correlation, though the small number of elements in the sample prevents deeper investigations. Taking into account only the Seyfert galaxies, the Pearson correlation coefficient emerges as $R=0.78$, meaning the correlation becomes stronger when we add the two blazars. We would like to emphasize that this plot shows for the first time the possibility that signatures from blazars might have the same origin as the Seyferts, which is a very different scientific point compared to \citet{Neronov2023}. We note, these six sources in our study are the best neutrino-source candidates at the present. We will initiate and analyze more hard X-ray observations first about those neutrino-source candidate blazars, for which we only know upper limits on the hard X-ray flux. This will help us to identify a control sample for our studies.

An important point is to understand the level of contribution of the jet component to the hard X-ray emission, especially in the case of \txs. We show below that using only those \nustar observations on \txs that were taken during the low very high-energy activity phase of \txs as measured by MAGIC \citep[$E>90$ GeV,][]{Ansoldi2018}, the resulted hard X-ray luminosity  
($(6.6\pm0.9)\times 10^{44}$\ergs) does not differ significantly from the mean one in Fig. \ref{fig:neronov-extended} ($(9.0\pm2.4)\times 10^{44}$\ergs, taking into account all \nustar observations). Adding to the picture that the \textit{Fermi}-LAT $\gamma$-ray emission goes down while the \textit{NuSTAR} hard X-ray emission goes up in the overlapping time period (see Fig. \ref{fig:fermi-nustar}), 
it is highly unlikely that the jet significantly contributes to the 15--55 keV hard X-rays of \txs (at least after 2017).

\subsection{Does the jet significantly contribute to the hard X-ray flux of \txs?}

\begin{figure}    
    \includegraphics[width=0.45\textwidth]{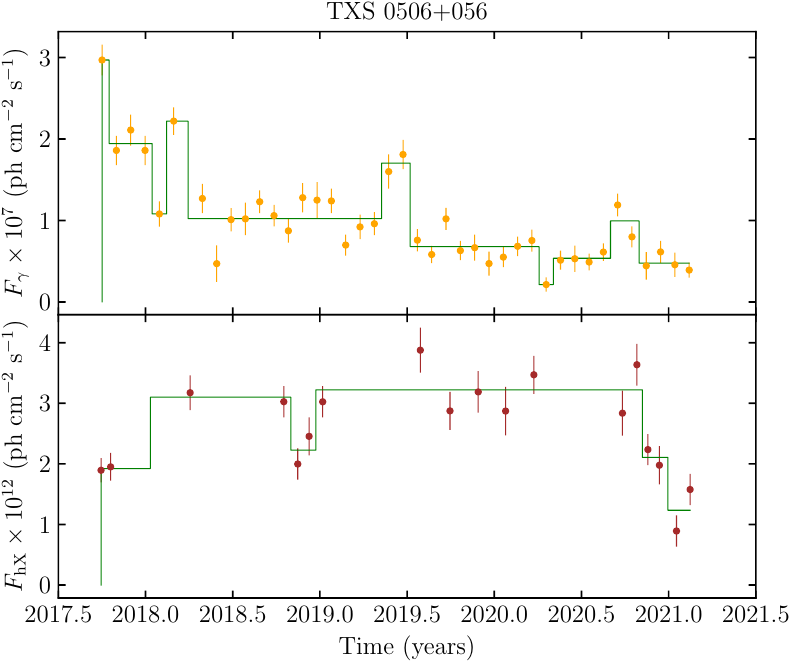}
    \caption{The upper panels shows the 0.1--100 GeV \textit{Fermi}-LAT one-month binned $\gamma$-ray light curve of \txs from the Fermi Light Curve Repository \citep{Abdollahi2023} (orange dots with error-bars) and the Bayesian blocks with $p=0.05$ (green line). The lower panels shows the 15--55 keV \textit{NuSTAR} hard X-ray light curve of \txs (red dots with error-bars) and the Bayesian blocks with $p=0.05$ (green line).}
    \label{fig:fermi-nustar}
\end{figure}

\citet{Ansoldi2018} shows the high-energy MAGIC observations ($E>90$~GeV) on \txs in about $40$~days from the detection time of the IceCube neutrino event IC-170922A. They identified two very-high energy (VHE) phases of \txs, one between MJD 58029–58030 and a second one on MJD 58057. 
The first two \nustar observations on \txs (2017-09-29 and 2017-10-19) overlapped with the MAGIC observations, and the MAGIC flux was low during the \nustar epochs. The mean \nustar luminosity in these two epochs is $(6.6\pm0.9)\times 10^{44}~\mathrm{erg}~\mathrm{s}^{-1}$, which does not differ significantly from the mean luminosity plotted in Fig. \ref{fig:neronov-extended} (which is $(9.0\pm2.4)\times 10^{44}~\mathrm{erg}~\mathrm{s}^{-1}$, corrected for variability, taking into account all \nustar observations). Another argument against a jet component dominating the hard X-ray emission shows up when comparing the 0.1-100 GeV \textit{Fermi}-LAT \citep{Abdollahi2023} and \nustar light curves (\nustar started to observe \txs only after the IC-170922A neutrino event). It seems that in the overlapping time period, the $\gamma$-ray emission, usually attributed to the jet component, goes down, while the hard X-ray emission goes up and maintains a high state, only dropping after the end of 2020 (see Fig. \ref{fig:fermi-nustar}). We note that the background of \txs is usually strong at lower energies.

Comparable unabsorbed hard X-ray and neutrino luminosities for blazars \txs and \gb, as well as four Seyfert galaxies suggests that neutrinos are produced in $\gamma$-obscured regions, such as their cores rather than along the jet, for all six sources. 

\subsection{Comparable hard X-ray and neutrino luminosity from $\gamma$-obscured sources}
\label{appendix:c}

Consider the case of protons accelerated near the black hole or in the accretion disk, interacting with a target photon field of characteristic size $\rm R_\mathrm{target}$ centered at the black hole. In this ``corona" region, which has large densities in both X-rays and accreting matter, the opacity to accelerated protons is \citep{2022IJMPD..3130003H}
\begin{equation}
\label{eq:pgammaopacity}
\rm \tau_{p\gamma}\simeq \frac{\kappa_{p\gamma} R_{target}}{\lambda_{p\gamma}} \simeq \kappa_{p\gamma} \rm R_{target}\, \sigma_{p\gamma} \,n_{\gamma},
\end{equation}
determined by how many times a proton interacts in a target of size $\rm R_{target}$ given its interaction length $\lambda_{p\gamma}$; $\kappa_{p\gamma}$ is inelasticity, or the fraction of the energy a proton loses with each interaction. The interaction length is determined by the density of target photons $\mathrm{n}_\gamma$ and the interaction cross section $\sigma_{p\gamma}$. For the simple dimensional analysis in this section we use the following cross sections $\rm \sigma_{\gamma\gamma} = 6.7 \times 10^{-25}~\mathrm{cm}^2$, $\rm \sigma_{p\gamma} = 5 \times 10^{-28}~\mathrm{cm}^2$, and $\rm \sigma_{pp} = 3 \times 10^{-26} \mathrm{cm}^2$.

The opacity of the target to the photons (pionic $\gamma$-rays) produced along with the neutrinos is given by
\begin{equation}
\tau_{\gamma\gamma} \simeq \rm R_{target}\, \sigma_{\gamma\gamma}\,n_{\gamma},
\end{equation}
and therefore, approximately, the two opacities are related by their cross sections 
\begin{equation}
\tau_{\gamma\gamma} \simeq \frac{\sigma_{\gamma\gamma}}{\kappa_{p\gamma} \sigma_{p\gamma}} \, \tau_{p\gamma} \simeq 10^3\,  \tau_{p\gamma}
\end{equation}
for $\rm R_{target} \sim R$, where $R$ is the size of the injection region. There is an additional factor associated with the different thresholds of the two interactions \citep{1987MNRAS.227..403S}. A target that produces neutrinos with $\tau_{p\gamma} \gtrsim 0.1$ will not be transparent to the pionic $\gamma$-rays, which will lose energy in the target even before propagating in the extragalactic background light. 

It has been shown by \citet{2022ApJ...941L..17M} that the linear scaling between $L_X$ and $L_\nu$ in Seyfert/low-luminosity (LL) AGN systems holds while $t^\star \approx R/V$ is longer than $R/c$, where $R$ is the emission radius, $V$ is the characteristic velocity in the plasma, and $c$ is the speed of light in the medium (see their Eq. 5 defining the photomeson production optical depth). For non-relativistic sources such as AGN coronae, the optical depth for the photomeson production can be larger by a factor of $10$--$100$.

\section{Common origin?}

AGNs can produce neutrinos without Earth-pointing relativistic jets. It is tempting to investigate whether a mechanism that does not require a jet pointing towards Earth, could be behind neutrino emission in blazars such as TXS~0506+056. 

One possible clue can be the $\sim1$\,MeV photon cutoff energy and the corresponding electromagnetic energy release in hard X-rays, which is natural in $\gamma$-ray attenuated AGN core environments but it is not justified in jets. For example, the observed hard X-ray flux during the neutrino-bright phase of TXS~0506+056 is well described by a lepto-hadronic model predicting a monotonically increasing flux to the GeV $\gamma$-ray band \citep{Walter1}. 
While current observations are not sensitive to probe the possible $\sim 1$\,MeV cutoff for TXS~0506+056, the comparable unabsorbed hard X-ray luminosity to that of the neutrinos supports a significant $\gamma$-ray energy release to the sub-MeV band. 

Second, blazars can produce detectable very high-energy (VHE, $> 100$ GeV) $\gamma$-rays \citep[e.g. the VHE-phases of \txs presented in][]{Ansoldi2018}. The escape of these VHE photons conflicts the apparent $\gamma$-obscured neutrino production that is needed by the comparable hard X-ray and neutrino fluxes. A possible reconciliation of VHE emission by blazar jets can be the temporal variability of $\gamma$ attenuation. When the jet is in a VHE-active phase, the Doppler-boosted jet outshines the corona/accretion disk, however, the same cannot happen for neutrinos. The temporal $\gamma$ suppression in jets is also consistent with the observed dip in $\gamma$ emission around neutrino observations reported for several blazars \citep{2021ApJ...911L..18K,Kun2022MW}. 

One additional clue may be the relative luminosities of the observed neutrino sources. TXS~0506+056 appears to be $\sim 2$ orders of magnitude brighter than the four Seyfert galaxies considered here (both in neutrinos and hard X-rays; see Fig.~\ref{fig:neronov-extended}), which might be internal or external to the core (e.g. jet viewing angle). Supermassive black holes (SMBHs) in blazars appear to have 1-2 orders of magnitude larger mass compared to the SMBHs in local galaxies. Since the luminosity scales with the total mass of the SMBH, this difference in the total masses naturally would resolve the difference in luminosities.

Another difference of note is the neutrino spectrum. The 2014--2015 burst of TXS~0506+056 appears to produce neutrino spectrum of $dN/dE\propto E_\nu^{-2.2}$ \citep{2018Sci...361..147I}. This is in contrast of the steeper spectra of, e.g., NGC 1068 which has a spectrum of $dN/dE\propto E_\nu^{-3.3}$ \citep{ngc1068_2022}. The reason is possibly that in the case of TXS~0506+056 we are observing high-energy neutrinos produced by $p\gamma$ interactions (e.g., \citet{1996tgra.conf..341S}), while somewhat lower energy neutrinos in NGC 1068 are produced via $pp$ interactions (e.g., \citet{2020PhRvL.125a1101M}).

The spectral index of the hard X-ray emission can hint at the underlying radiative processes. The spectral indices of the Seyfert galaxies are:
\begin{itemize}
    \item NGC 1068: $\Gamma = 2.10 \pm 0.07$ \citep[Table 2. in][]{2015ApJ...812..116B}, but also $\Gamma = 1.73 \pm 0.07$ \citep[Table. 2 in ][]{Marchesi2018},
    \item NGC 3079: $\Gamma = 1.94 \pm 0.1$ \citep[Table 2. in ][]{Marchesi2018},
    \item NGC 4151: $\Gamma = 1.9 \pm 0.1$ \citep[Table 8. in][]{2015ApJ...806..149K},
    \item CGCG 420-015: $\Gamma=1.66^{+0.11}_{-0.12}$ \citep[Table 2. in ][]{Marchesi2018},
\end{itemize}
and similarly for the blazars:
\begin{itemize}
    \item \txs: $\Gamma = 1.5 - 1.9$  (our Section \ref{subsec:txs_nustar}; previous reports are $\Gamma \sim 1.69 \pm 0.13$),
    \item \gb: $\Gamma = 1.55 \pm  0.15$ (our Section \ref{subsec:gb6_nustar}).
\end{itemize}

We highlight the large uncertainties in these values. In some cases the spectral indices have large error-bars, whereas in NGC~1068 the values depend on the observations and analysis used. Moreover, in \txs there is also evidence of  variability in the spectral index. We also caution that not all spectral indices are derived with a consistent method, as the energy range in which the sources are detected is not always the same, plus some sources require a more complex model to derive the spectral index due to obscuration effects (particularly important in the case of NGC~1068). Thus, we cannot draw strong conclusions regarding the emission mechanisms. We can only conclude that the blazars seem to have slightly harder spectra, but still consistent with the values found in some of the Seyferts, compatible with an interpretation of a common origin of their X-ray emission.

Finally, the emission's temporal variation of multi-messenger events may help differentiate between a jet and non-jet origin. For example, neutrinos were detected from TXS~0506+056 in effectively two distinct emission episodes, in 2014/2015 and 2017 \citep{2018Sci...361..147I}. While the 2017 detection coincided with a several months-long $\gamma$ flare, TXS~0506+056 had low $E_\gamma > 100$\,MeV flux during the 2015 emission episode \citep{2018Sci...361..147I}. Radio flux was roughly constant throughout of the 2014/2015 episode, suggesting that changes in the relativistic outflow by itself do not account for the varying neutrino flux, rather the energy excess was suppressed already in the corona and did not reach the radio jet base. 

\citet{2024MNRAS.529L.135G} pointed out that winds and jets all carry an electric current. If the jets or winds vary with time, and they  all do, then the temporal variation of the electric currents produces temporary electric fields. These fields discharge, producing energetic particles with spectra between $p^{-2}$ and $p^{-4}$ for hadrons and $p^{-3}$ and $p^{-5}$ for electrons/positrons, in the Kardashev \citep{1962SvA.....6..317K} loss limit. This gives a spectral range in synchrotron emission in the range
between ${\nu}^{-1}$ and ${\nu}^{-2}$, which is widely observed in radio filaments, both Galactic and extragalactic. 

\section{Conclusion}

We found that the unabsorbed hard X-ray and high-energy neutrino luminosities of blazar TXS~0506+056 are comparable. This is similar to the relation for Seyfert galaxies NGC~1068, NGC~4151, \cgcg and NGC~3079 (see left panel in Fig.~\ref{fig:neronov-extended}). Another blazar, \gb also seems to be close to the relation. This relation for Seyfert galaxies was initially found by \cite{Neronov2023}, but here we recalculate it using hard X-ray flux measurements from \textit{NuSTAR} observations in the literature. We would like to emphasize that we did not use a broad-band spectral modeling to suggest the corona dominates the hard X-ray emission of \txs in the energy range 15-55 keV, instead of this we rely on observations. The fact that the hard X-ray -- neutrino flux ratio for the two blazars fits that of the Seyferts hints at that similar emission zones dominates in blazars, at least in that energy range that is just above the spectral hardening above 10 keV, seen for e.g. NGC 1068 \cite{ngc1068_2022} and \txs \cite{ICTXS2018a}. We hypothesize that the correlation between hard X-rays and neutrino fluxes is due to a common origin in the corona. However, a larger sample is required to test this hypothesis, or whether alternative scenarios are favored.

Our results suggest the following:
\begin{enumerate}[noitemsep]
\item The comparable unabsorbed hard X-ray and neutrino luminosities in our sample is consistent with neutrino production in $\gamma$-obscure regions with photons attenuated down to $\sim 1$\,MeV energies. 
\item The same astrophysical process might be responsible for neutrino production in blazars and Seyfert AGNs. Photon attenuation to $\sim 1$\,MeV is expected for neutrino production near AGN disks. In this case, neutrino production may not be dominated by jets, even in the case of blazars.
\item Hard X-rays are very promising targets for multi-messenger modeling of AGN in respect of neutrino-source searches.
\end{enumerate}
We note that all of the potential classes of neutrino sources have the same astrophysical nature: accreting supermassive black holes. Since neutrino emission from Seyfert galaxies is unlikely to be related to their weak jets, a natural question arises as to whether neutrino emission from blazars could also have an origin besides their powerful jets, such as near the central black hole.

We caution nonetheless that the comparable X-ray and neutrino fluxes we found are both subject to uncertainties due to, e.g., emission's spectral uncertainty and possible temporal variations. Challenges include the generally weak and not well time-constrained neutrino signal versus X-ray observation window, the more sophisticated determination of the unabsorbed hard X-ray flux, sample size. It will be particularly interesting to determine whether a similar relation holds for other identified sources of high-energy neutrinos, and whether multi-messenger emission and spectral features can be used to distinguish between the disk/corona and jet origins of neutrinos. We encourage deep hard X-ray/soft-$\gamma$-ray observations of these sources.

\begin{acknowledgments}
The authors thank the Referees for the thoughtful and constructive points. The authors also would like to thank Björn Eichmann, Stefano Marchesi, Sera Markoff, Kohta Murase, Claudio Ricci, Walter Winter and Xiurui Zhao for valuable discussions. E.K. thanks the Alexander von Humboldt Foundation for its Fellowship. J.B.T.\, E.K., and A.F.\ acknowledge support from the German Science Foundation DFG, via the Collaborative Research Center \textit{SFB1491: Cosmic Interacting Matters -- from Source to Signal} (grant no.\ 445052434). S.d.P. acknowledges support from ERC Advanced Grant 789410. This research has made use of the NuSTAR Data Analysis Software (NuSTARDAS) jointly developed by the ASI Space Science Data Center (SSDC, Italy) and the California Institute of Technology (Caltech, USA). This work made use of data supplied by the UK Swift Science Data Centre at the University of Leicester. This work made use of Astropy: \url{http://www.astropy.org}, a community-developed core Python package and an ecosystem of tools and resources for astronomy \cite{astropy:2013, astropy:2022}.
\end{acknowledgments}


\begin{thebibliography}{61}%
\makeatletter
\providecommand \@ifxundefined [1]{%
 \@ifx{#1\undefined}
}%
\providecommand \@ifnum [1]{%
 \ifnum #1\expandafter \@firstoftwo
 \else \expandafter \@secondoftwo
 \fi
}%
\providecommand \@ifx [1]{%
 \ifx #1\expandafter \@firstoftwo
 \else \expandafter \@secondoftwo
 \fi
}%
\providecommand \natexlab [1]{#1}%
\providecommand \enquote  [1]{``#1''}%
\providecommand \bibnamefont  [1]{#1}%
\providecommand \bibfnamefont [1]{#1}%
\providecommand \citenamefont [1]{#1}%
\providecommand \href@noop [0]{\@secondoftwo}%
\providecommand \href [0]{\begingroup \@sanitize@url \@href}%
\providecommand \@href[1]{\@@startlink{#1}\@@href}%
\providecommand \@@href[1]{\endgroup#1\@@endlink}%
\providecommand \@sanitize@url [0]{\catcode `\\12\catcode `\$12\catcode `\&12\catcode `\#12\catcode `\^12\catcode `\_12\catcode `\%12\relax}%
\providecommand \@@startlink[1]{}%
\providecommand \@@endlink[0]{}%
\providecommand \url  [0]{\begingroup\@sanitize@url \@url }%
\providecommand \@url [1]{\endgroup\@href {#1}{\urlprefix }}%
\providecommand \urlprefix  [0]{URL }%
\providecommand \Eprint [0]{\href }%
\providecommand \doibase [0]{https://doi.org/}%
\providecommand \selectlanguage [0]{\@gobble}%
\providecommand \bibinfo  [0]{\@secondoftwo}%
\providecommand \bibfield  [0]{\@secondoftwo}%
\providecommand \translation [1]{[#1]}%
\providecommand \BibitemOpen [0]{}%
\providecommand \bibitemStop [0]{}%
\providecommand \bibitemNoStop [0]{.\EOS\space}%
\providecommand \EOS [0]{\spacefactor3000\relax}%
\providecommand \BibitemShut  [1]{\csname bibitem#1\endcsname}%
\let\auto@bib@innerbib\@empty
\bibitem [{\citenamefont {{Aartsen}}\ \emph {et~al.}(2013)\citenamefont {{Aartsen}}, \citenamefont {{Abbasi}}, \citenamefont {{Abdou}}, \citenamefont {{Ackermann}}, \citenamefont {{Adams}} \emph {et~al.}}]{2013PhRvL.111b1103A}%
  \BibitemOpen
  \bibfield  {author} {\bibinfo {author} {\bibfnamefont {M.~G.}\ \bibnamefont {{Aartsen}}}, \bibinfo {author} {\bibfnamefont {R.}~\bibnamefont {{Abbasi}}}, \bibinfo {author} {\bibfnamefont {Y.}~\bibnamefont {{Abdou}}}, \bibinfo {author} {\bibfnamefont {M.}~\bibnamefont {{Ackermann}}}, \bibinfo {author} {\bibfnamefont {J.}~\bibnamefont {{Adams}}}, \emph {et~al.},\ }\href {https://doi.org/10.1103/PhysRevLett.111.021103} {\bibfield  {journal} {\bibinfo  {journal} {\prl}\ }\textbf {\bibinfo {volume} {111}},\ \bibinfo {eid} {021103} (\bibinfo {year} {2013})},\ \Eprint {https://arxiv.org/abs/1304.5356} {arXiv:1304.5356 [astro-ph.HE]} \BibitemShut {NoStop}%
\bibitem [{\citenamefont {{IceCube Collaboration}}\ \emph {et~al.}(2018)\citenamefont {{IceCube Collaboration}}, \citenamefont {{Fermi-LAT}}, \citenamefont {{MAGIC}}, \citenamefont {{AGILE}}, \citenamefont {{ASAS-SN}}, \citenamefont {{HAWC}}, \citenamefont {{H.E.S.S.}}, \citenamefont {{INTEGRAL}}, \citenamefont {{Kanata}}, \citenamefont {{Kiso}}, \citenamefont {{Kapteyn}}, \citenamefont {{Liverpool Telescope}}, \citenamefont {{Subaru}}, \citenamefont {{Swift/NuSTAR}}, \citenamefont {{VERITAS}},\ and\ \citenamefont {{VLA/17B-403}}}]{ICTXS2018a}%
  \BibitemOpen
  \bibfield  {author} {\bibinfo {author} {\bibnamefont {{IceCube Collaboration}}}, \bibinfo {author} {\bibnamefont {{Fermi-LAT}}}, \bibinfo {author} {\bibnamefont {{MAGIC}}}, \bibinfo {author} {\bibnamefont {{AGILE}}}, \bibinfo {author} {\bibnamefont {{ASAS-SN}}}, \bibinfo {author} {\bibnamefont {{HAWC}}}, \bibinfo {author} {\bibnamefont {{H.E.S.S.}}}, \bibinfo {author} {\bibnamefont {{INTEGRAL}}}, \bibinfo {author} {\bibnamefont {{Kanata}}}, \bibinfo {author} {\bibnamefont {{Kiso}}}, \bibinfo {author} {\bibnamefont {{Kapteyn}}}, \bibinfo {author} {\bibnamefont {{Liverpool Telescope}}}, \bibinfo {author} {\bibnamefont {{Subaru}}}, \bibinfo {author} {\bibnamefont {{Swift/NuSTAR}}}, \bibinfo {author} {\bibnamefont {{VERITAS}}},\ and\ \bibinfo {author} {\bibnamefont {{VLA/17B-403}}},\ }\href {https://doi.org/10.1126/science.aat1378} {\bibfield  {journal} {\bibinfo  {journal} {Science}\ }\textbf {\bibinfo {volume} {361}},\ \bibinfo {pages} {147} (\bibinfo {year} {2018})}\BibitemShut {NoStop}%
\bibitem [{\citenamefont {{Mannheim}}\ \emph {et~al.}(1992)\citenamefont {{Mannheim}}, \citenamefont {{Stanev}},\ and\ \citenamefont {{Biermann}}}]{msb1992}%
  \BibitemOpen
  \bibfield  {author} {\bibinfo {author} {\bibfnamefont {K.}~\bibnamefont {{Mannheim}}}, \bibinfo {author} {\bibfnamefont {T.}~\bibnamefont {{Stanev}}},\ and\ \bibinfo {author} {\bibfnamefont {P.~L.}\ \bibnamefont {{Biermann}}},\ }\href@noop {} {\bibfield  {journal} {\bibinfo  {journal} {\aap}\ }\textbf {\bibinfo {volume} {260}},\ \bibinfo {pages} {L1} (\bibinfo {year} {1992})}\BibitemShut {NoStop}%
\bibitem [{\citenamefont {{Becker}}\ \emph {et~al.}(2005)\citenamefont {{Becker}}, \citenamefont {{Biermann}},\ and\ \citenamefont {{Rhode}}}]{bbr2005}%
  \BibitemOpen
  \bibfield  {author} {\bibinfo {author} {\bibfnamefont {J.~K.}\ \bibnamefont {{Becker}}}, \bibinfo {author} {\bibfnamefont {P.~L.}\ \bibnamefont {{Biermann}}},\ and\ \bibinfo {author} {\bibfnamefont {W.}~\bibnamefont {{Rhode}}},\ }\href {https://doi.org/10.1016/j.astropartphys.2005.02.003} {\bibfield  {journal} {\bibinfo  {journal} {Astroparticle Physics}\ }\textbf {\bibinfo {volume} {23}},\ \bibinfo {pages} {355} (\bibinfo {year} {2005})},\ \Eprint {https://arxiv.org/abs/astro-ph/0502089} {arXiv:astro-ph/0502089 [astro-ph]} \BibitemShut {NoStop}%
\bibitem [{\citenamefont {{Becker}}(2008)}]{physirep2008}%
  \BibitemOpen
  \bibfield  {author} {\bibinfo {author} {\bibfnamefont {J.~K.}\ \bibnamefont {{Becker}}},\ }\href {https://doi.org/10.1016/j.physrep.2007.10.006} {\bibfield  {journal} {\bibinfo  {journal} {\physrep}\ }\textbf {\bibinfo {volume} {458}},\ \bibinfo {pages} {173} (\bibinfo {year} {2008})},\ \Eprint {https://arxiv.org/abs/0710.1557} {arXiv:0710.1557 [astro-ph]} \BibitemShut {NoStop}%
\bibitem [{\citenamefont {{Becker Tjus}}\ \emph {et~al.}(2014)\citenamefont {{Becker Tjus}}, \citenamefont {{Eichmann}}, \citenamefont {{Halzen}}, \citenamefont {{Kheirandish}},\ and\ \citenamefont {{Saba}}}]{radiogalaxies2014}%
  \BibitemOpen
  \bibfield  {author} {\bibinfo {author} {\bibfnamefont {J.}~\bibnamefont {{Becker Tjus}}}, \bibinfo {author} {\bibfnamefont {B.}~\bibnamefont {{Eichmann}}}, \bibinfo {author} {\bibfnamefont {F.}~\bibnamefont {{Halzen}}}, \bibinfo {author} {\bibfnamefont {A.}~\bibnamefont {{Kheirandish}}},\ and\ \bibinfo {author} {\bibfnamefont {S.~M.}\ \bibnamefont {{Saba}}},\ }\href {https://doi.org/10.1103/PhysRevD.89.123005} {\bibfield  {journal} {\bibinfo  {journal} {\prd}\ }\textbf {\bibinfo {volume} {89}},\ \bibinfo {eid} {123005} (\bibinfo {year} {2014})},\ \Eprint {https://arxiv.org/abs/1406.0506} {arXiv:1406.0506 [astro-ph.HE]} \BibitemShut {NoStop}%
\bibitem [{\citenamefont {{Araudo}}\ \emph {et~al.}(2010)\citenamefont {{Araudo}}, \citenamefont {{Bosch-Ramon}},\ and\ \citenamefont {{Romero}}}]{2010A&A...522A..97A}%
  \BibitemOpen
  \bibfield  {author} {\bibinfo {author} {\bibfnamefont {A.~T.}\ \bibnamefont {{Araudo}}}, \bibinfo {author} {\bibfnamefont {V.}~\bibnamefont {{Bosch-Ramon}}},\ and\ \bibinfo {author} {\bibfnamefont {G.~E.}\ \bibnamefont {{Romero}}},\ }\href {https://doi.org/10.1051/0004-6361/201014660} {\bibfield  {journal} {\bibinfo  {journal} {\aap}\ }\textbf {\bibinfo {volume} {522}},\ \bibinfo {eid} {A97} (\bibinfo {year} {2010})},\ \Eprint {https://arxiv.org/abs/1007.2199} {arXiv:1007.2199 [astro-ph.HE]} \BibitemShut {NoStop}%
\bibitem [{\citenamefont {{Tavecchio}}\ and\ \citenamefont {{Ghisellini}}(2015)}]{2015MNRAS.451.1502T}%
  \BibitemOpen
  \bibfield  {author} {\bibinfo {author} {\bibfnamefont {F.}~\bibnamefont {{Tavecchio}}}\ and\ \bibinfo {author} {\bibfnamefont {G.}~\bibnamefont {{Ghisellini}}},\ }\href {https://doi.org/10.1093/mnras/stv1023} {\bibfield  {journal} {\bibinfo  {journal} {\mnras}\ }\textbf {\bibinfo {volume} {451}},\ \bibinfo {pages} {1502} (\bibinfo {year} {2015})},\ \Eprint {https://arxiv.org/abs/1411.2783} {arXiv:1411.2783 [astro-ph.HE]} \BibitemShut {NoStop}%
\bibitem [{\citenamefont {{Abbasi}}\ \emph {et~al.}(2022{\natexlab{a}})\citenamefont {{Abbasi}}, \citenamefont {{Ackermann}}, \citenamefont {{Adams}}, \citenamefont {{Aguilar}}, \citenamefont {{Ahlers}}, \citenamefont {{Ahrens}}, \citenamefont {{Alameddine}}, \citenamefont {{Alispach}}  \emph {et~al.}}]{ngc1068_2022}%
  \BibitemOpen
  \bibfield  {author} {\bibinfo {author} {\bibfnamefont {R.}~\bibnamefont {{Abbasi}}}, \bibinfo {author} {\bibfnamefont {M.}~\bibnamefont {{Ackermann}}}, \bibinfo {author} {\bibfnamefont {J.}~\bibnamefont {{Adams}}}, \bibinfo {author} {\bibfnamefont {J.~A.}\ \bibnamefont {{Aguilar}}}, \bibinfo {author} {\bibfnamefont {M.}~\bibnamefont {{Ahlers}}}, \bibinfo {author} {\bibfnamefont {M.}~\bibnamefont {{Ahrens}}}, \bibinfo {author} {\bibfnamefont {J.~M.}\ \bibnamefont {{Alameddine}}}, \bibinfo {author} {\bibfnamefont {C.}~\bibnamefont {{Alispach}}}, \emph {et~al.},\ }\href {https://doi.org/10.1126/science.abg3395} {\bibfield  {journal} {\bibinfo  {journal} {Science}\ }\textbf {\bibinfo {volume} {378}},\ \bibinfo {pages} {538} (\bibinfo {year} {2022}{\natexlab{a}})},\ \Eprint {https://arxiv.org/abs/2211.09972} {arXiv:2211.09972 [astro-ph.HE]} \BibitemShut {NoStop}%
\bibitem [{\citenamefont {{Taboada}}\ and\ \citenamefont {{Stein}}(2019)}]{2019ATel12967....1T}%
  \BibitemOpen
  \bibfield  {author} {\bibinfo {author} {\bibfnamefont {I.}~\bibnamefont {{Taboada}}}\ and\ \bibinfo {author} {\bibfnamefont {R.}~\bibnamefont {{Stein}}},\ }\href@noop {} {\bibfield  {journal} {\bibinfo  {journal} {The Astronomer's Telegram}\ }\textbf {\bibinfo {volume} {12967}},\ \bibinfo {pages} {1} (\bibinfo {year} {2019})}\BibitemShut {NoStop}%
\bibitem [{\citenamefont {{Kadler}}\ \emph {et~al.}(2016)\citenamefont {{Kadler}}, \citenamefont {{Krau{\ss}}}, \citenamefont {{Mannheim}}, \citenamefont {{Ojha}}, \citenamefont {{M{\"u}ller}}, \citenamefont {{Schulz}}, \citenamefont {{Anton}}, \citenamefont {{Baumgartner}}, \emph {et~al.},\ }]{2016NatPh..12..807K}%
  \BibitemOpen
  \bibfield  {author} {\bibinfo {author} {\bibfnamefont {M.}~\bibnamefont {{Kadler}}}, \bibinfo {author} {\bibfnamefont {F.}~\bibnamefont {{Krau{\ss}}}}, \bibinfo {author} {\bibfnamefont {K.}~\bibnamefont {{Mannheim}}}, \bibinfo {author} {\bibfnamefont {R.}~\bibnamefont {{Ojha}}}, \bibinfo {author} {\bibfnamefont {C.}~\bibnamefont {{M{\"u}ller}}}, \bibinfo {author} {\bibfnamefont {R.}~\bibnamefont {{Schulz}}}, \bibinfo {author} {\bibfnamefont {G.}~\bibnamefont {{Anton}}}, \bibinfo {author} {\bibfnamefont {W.}~\bibnamefont {{Baumgartner}}}, \emph {et~al.},\ }\href {https://doi.org/10.1038/nphys3715} {\bibfield  {journal} {\bibinfo  {journal} {Nature Physics}\ }\textbf {\bibinfo {volume} {12}},\ \bibinfo {pages} {807} (\bibinfo {year} {2016})},\ \Eprint {https://arxiv.org/abs/1602.02012} {arXiv:1602.02012 [astro-ph.HE]} \BibitemShut {NoStop}%
\bibitem [{\citenamefont {{Goswami}}(2023)}]{2023arXiv230715349G}%
  \BibitemOpen
  \bibfield  {author} {\bibinfo {author} {\bibfnamefont {S.}~\bibnamefont {{Goswami}}},\ }\href {https://doi.org/10.48550/arXiv.2307.15349} {\bibfield  {journal} {\bibinfo  {journal} {arXiv e-prints}\ ,\ \bibinfo {eid} {arXiv:2307.15349}} (\bibinfo {year} {2023})},\ \Eprint {https://arxiv.org/abs/2307.15349} {arXiv:2307.15349 [astro-ph.HE]} \BibitemShut {NoStop}%
\bibitem [{\citenamefont {{Neronov}}\ \emph {et~al.}(2024)\citenamefont {{Neronov}}, \citenamefont {{Savchenko}},\ and\ \citenamefont {{Semikoz}}}]{Neronov2023}%
  \BibitemOpen
  \bibfield  {author} {\bibinfo {author} {\bibfnamefont {A.}~\bibnamefont {{Neronov}}}, \bibinfo {author} {\bibfnamefont {D.}~\bibnamefont {{Savchenko}}},\ and\ \bibinfo {author} {\bibfnamefont {D.~V.}\ \bibnamefont {{Semikoz}}},\ }\href {https://doi.org/ 
10.1103/PhysRevLett.132.101002} {\bibfield  {journal} {\bibinfo  {journal} {\prl}\ }\textbf {\bibinfo {volume} {132}},\ \bibinfo {eid} {101002} (\bibinfo {year} {2024})} \BibitemShut {NoStop}%
  \bibitem [{\citenamefont {{Abbasi}}\ \emph {et~al.}(2024{\natexlab{a}})\citenamefont {{Abbasi}}, \citenamefont {{Ackermann}}, \citenamefont {{Adams}}, \citenamefont {{Aguilar}}, \citenamefont {{Ahlers}}, \citenamefont {{Ahrens}}, \citenamefont {{Alameddine}}, \citenamefont {{Alispach}}  \emph {et~al.}}]{Abassi2024}%
  \BibitemOpen
  \bibfield  {author} {\bibinfo {author} {\bibfnamefont {R.}~\bibnamefont {{Abbasi}}}, \bibinfo {author} {\bibfnamefont {M.}~\bibnamefont {{Ackermann}}}, \bibinfo {author} {\bibfnamefont {J.}~\bibnamefont {{Adams}}}, \bibinfo {author} {\bibfnamefont {J.~A.}\ \bibnamefont {{Aguilar}}}, \bibinfo {author} {\bibfnamefont {M.}~\bibnamefont {{Ahlers}}}, \bibinfo {author} {\bibfnamefont {M.}~\bibnamefont {{Ahrens}}}, \bibinfo {author} {\bibfnamefont {J.~M.}\ \bibnamefont {{Alameddine}}}, \bibinfo {author} {\bibfnamefont {C.}~\bibnamefont {{Alispach}}}, \emph {et~al.},\ }\href {https://doi.org/10.1126/10.48550/arXiv.2406.07601} \Eprint {https://arxiv.org/abs/arXiv:2406.07601} {arXiv: arXiv:2406.07601 [astro-ph.HE]} \BibitemShut {NoStop}%
   \bibitem [{\citenamefont {{Padovani}}\ \emph {et~al.}(2022)\citenamefont {{Padovani}}, \citenamefont {{Boccardi}}, \citenamefont {{Falomo}}\ and\ \citenamefont {{Giommi}}}]{Padovani2022}%
  \BibitemOpen
  \bibfield  {author} {\bibinfo {author} {\bibfnamefont {P.}~\bibnamefont {{Padovani}}}, \bibinfo {author} {\bibfnamefont {B.}~\bibnamefont {{Boccardi}}}, \bibinfo {author} {\bibfnamefont {R.}~\bibnamefont {{Falomo}}},\ and\ \bibinfo {author} {\bibfnamefont {P.}\ \bibnamefont {{Giommi}}},\ }\href {https://doi.org/ 
10.1093/mnras/stac376} {\bibfield  {journal} {\bibinfo  {journal} {\mnras}\ }\textbf {\bibinfo {volume} {511}},\ \bibinfo {eid} {4697} (\bibinfo {year} {2022})},\ \Eprint {https://arxiv.org/abs/2202.04363} {arXiv:2202.04363 [astro-ph.HE]} \BibitemShut {NoStop}%
\bibitem [{\citenamefont {{Murase}}\ \emph {et~al.}(2020{\natexlab{a}})\citenamefont {{Murase}}, \citenamefont {{Kimura}},\ and\ \citenamefont {{M{\'e}sz{\'a}ros}}}]{2020PhRvL.125a1101M}%
  \BibitemOpen
  \bibfield  {author} {\bibinfo {author} {\bibfnamefont {K.}~\bibnamefont {{Murase}}}, \bibinfo {author} {\bibfnamefont {S.~S.}\ \bibnamefont {{Kimura}}},\ and\ \bibinfo {author} {\bibfnamefont {P.}~\bibnamefont {{M{\'e}sz{\'a}ros}}},\ }\href {https://doi.org/10.1103/PhysRevLett.125.011101} {\bibfield  {journal} {\bibinfo  {journal} {\prl}\ }\textbf {\bibinfo {volume} {125}},\ \bibinfo {eid} {011101} (\bibinfo {year} {2020}{\natexlab{a}})},\ \Eprint {https://arxiv.org/abs/1904.04226} {arXiv:1904.04226 [astro-ph.HE]} \BibitemShut {NoStop}%
\bibitem [{\citenamefont {{Inoue}}\ \emph {et~al.}(2020)\citenamefont {{Inoue}}, \citenamefont {{Khangulyan}},\ and\ \citenamefont {{Doi}}}]{2020ApJ...891L..33I}%
  \BibitemOpen
  \bibfield  {author} {\bibinfo {author} {\bibfnamefont {Y.}~\bibnamefont {{Inoue}}}, \bibinfo {author} {\bibfnamefont {D.}~\bibnamefont {{Khangulyan}}},\ and\ \bibinfo {author} {\bibfnamefont {A.}~\bibnamefont {{Doi}}},\ }\href {https://doi.org/10.3847/2041-8213/ab7661} {\bibfield  {journal} {\bibinfo  {journal} {\apjl}\ }\textbf {\bibinfo {volume} {891}},\ \bibinfo {eid} {L33} (\bibinfo {year} {2020})},\ \Eprint {https://arxiv.org/abs/1909.02239} {arXiv:1909.02239 [astro-ph.HE]} \BibitemShut {NoStop}%
\bibitem [{\citenamefont {{Eichmann}}\ \emph {et~al.}(2022)\citenamefont {{Eichmann}}, \citenamefont {{Oikonomou}}, \citenamefont {{Salvatore}}, \citenamefont {{Dettmar}},\ and\ \citenamefont {{Becker Tjus}}}]{2022ApJ...939...43E}%
  \BibitemOpen
  \bibfield  {author} {\bibinfo {author} {\bibfnamefont {B.}~\bibnamefont {{Eichmann}}}, \bibinfo {author} {\bibfnamefont {F.}~\bibnamefont {{Oikonomou}}}, \bibinfo {author} {\bibfnamefont {S.}~\bibnamefont {{Salvatore}}}, \bibinfo {author} {\bibfnamefont {R.-J.}\ \bibnamefont {{Dettmar}}},\ and\ \bibinfo {author} {\bibfnamefont {J.}~\bibnamefont {{Becker Tjus}}},\ }\href {https://doi.org/10.3847/1538-4357/ac9588} {\bibfield  {journal} {\bibinfo  {journal} {\apj}\ }\textbf {\bibinfo {volume} {939}},\ \bibinfo {eid} {43} (\bibinfo {year} {2022})},\ \Eprint {https://arxiv.org/abs/2207.00102} {arXiv:2207.00102 [astro-ph.HE]} \BibitemShut {NoStop}%
\bibitem [{\citenamefont {{Kun}}\ \emph {et~al.}(2023)\citenamefont {{Kun}}, \citenamefont {{Bartos}}, \citenamefont {{Becker Tjus}}, \citenamefont {{Biermann}}, \citenamefont {{Franckowiak}}, \citenamefont {{Halzen}},\ and\ \citenamefont {{Mez{\H{o}}}}}]{Kun2023}%
  \BibitemOpen
  \bibfield  {author} {\bibinfo {author} {\bibfnamefont {E.}~\bibnamefont {{Kun}}}, \bibinfo {author} {\bibfnamefont {I.}~\bibnamefont {{Bartos}}}, \bibinfo {author} {\bibfnamefont {J.}~\bibnamefont {{Becker Tjus}}}, \bibinfo {author} {\bibfnamefont {P.~L.}\ \bibnamefont {{Biermann}}},  \bibinfo {author} {\bibfnamefont {A.}\ \bibnamefont {{Franckowiak}}},\bibinfo {author} {\bibfnamefont {F.}~\bibnamefont {{Halzen}}},\ and\ \bibinfo {author} {\bibfnamefont {G.}~\bibnamefont {{Mez{\H{o}}}}},\ }\href {https://doi.org/ 10.1051/0004-6361/202346710} {\bibfield  {journal} {\bibinfo  {journal} {\aap}\ }\textbf {\bibinfo {volume} {679}},\ \bibinfo {eid} {46} (\bibinfo {year} {2023})},\ \Eprint {https://arxiv.org/abs/2305.06729} {arXiv:2305.06729 [astro-ph.HE]} \BibitemShut {NoStop}%
\bibitem [{\citenamefont {{Kun}}\ \emph {et~al.}(2021)\citenamefont {{Kun}}, \citenamefont {{Bartos}}, \citenamefont {{Becker Tjus}}, \citenamefont {{Biermann}}, \citenamefont {{Halzen}},\ and\ \citenamefont {{Mez{\H{o}}}}}]{2021ApJ...911L..18K}%
  \BibitemOpen
  \bibfield  {author} {\bibinfo {author} {\bibfnamefont {E.}~\bibnamefont {{Kun}}}, \bibinfo {author} {\bibfnamefont {I.}~\bibnamefont {{Bartos}}}, \bibinfo {author} {\bibfnamefont {J.}~\bibnamefont {{Becker Tjus}}}, \bibinfo {author} {\bibfnamefont {P.~L.}\ \bibnamefont {{Biermann}}}, \bibinfo {author} {\bibfnamefont {F.}~\bibnamefont {{Halzen}}},\ and\ \bibinfo {author} {\bibfnamefont {G.}~\bibnamefont {{Mez{\H{o}}}}},\ }\href {https://doi.org/10.3847/2041-8213/abf1ec} {\bibfield  {journal} {\bibinfo  {journal} {\apjl}\ }\textbf {\bibinfo {volume} {911}},\ \bibinfo {eid} {L18} (\bibinfo {year} {2021})},\ \Eprint {https://arxiv.org/abs/2009.09792} {arXiv:2009.09792 [astro-ph.HE]} \BibitemShut {NoStop}%
\bibitem [{\citenamefont {Murase}\ \emph {et~al.}(2016)\citenamefont {Murase}, \citenamefont {Guetta},\ and\ \citenamefont {Ahlers}}]{PhysRevLett.116.071101}%
  \BibitemOpen
  \bibfield  {author} {\bibinfo {author} {\bibfnamefont {K.}~\bibnamefont {Murase}}, \bibinfo {author} {\bibfnamefont {D.}~\bibnamefont {Guetta}},\ and\ \bibinfo {author} {\bibfnamefont {M.}~\bibnamefont {Ahlers}},\ }\href {https://doi.org/10.1103/PhysRevLett.116.071101} {\bibfield  {journal} {\bibinfo  {journal} {Phys. Rev. Lett.}\ }\textbf {\bibinfo {volume} {116}},\ \bibinfo {pages} {071101} (\bibinfo {year} {2016})}\BibitemShut {NoStop}%
\bibitem [{\citenamefont {{Murase}}\ and\ \citenamefont {{Waxman}}(2016)}]{2016PhRvD..94j3006M}%
  \BibitemOpen
  \bibfield  {author} {\bibinfo {author} {\bibfnamefont {K.}~\bibnamefont {{Murase}}}\ and\ \bibinfo {author} {\bibfnamefont {E.}~\bibnamefont {{Waxman}}},\ }\href {https://doi.org/10.1103/PhysRevD.94.103006} {\bibfield  {journal} {\bibinfo  {journal} {\prd}\ }\textbf {\bibinfo {volume} {94}},\ \bibinfo {eid} {103006} (\bibinfo {year} {2016})},\ \Eprint {https://arxiv.org/abs/1607.01601} {arXiv:1607.01601 [astro-ph.HE]} \BibitemShut {NoStop}%
\bibitem [{\citenamefont {{Meyer}}\ \emph {et~al.}(2004)\citenamefont {{Meyer}}, \citenamefont {{Zwaan}}, \citenamefont {{Webster}}, \citenamefont {{Staveley-Smith}}, \citenamefont {{Ryan-Weber}}, \citenamefont {{Drinkwater}}, \citenamefont {{Barnes}}, \citenamefont {{Howlett}}, \citenamefont {{Kilborn}}, \citenamefont {{Stevens}}, \citenamefont {{Waugh}}, \citenamefont {{Pierce}}, \citenamefont {{Bhathal}}, \citenamefont {{de Blok}}, \citenamefont {{Disney}}, \citenamefont {{Ekers}}, \citenamefont {{Freeman}}, \citenamefont {{Garcia}}, \citenamefont {{Gibson}}, \citenamefont {{Harnett}}, \citenamefont {{Henning}}, \citenamefont {{Jerjen}}, \citenamefont {{Kesteven}}, \citenamefont {{Knezek}}, \citenamefont {{Koribalski}}, \citenamefont {{Mader}}, \citenamefont {{Marquarding}}, \citenamefont {{Minchin}}, \citenamefont {{O'Brien}}, \citenamefont {{Oosterloo}}, \citenamefont {{Price}}, \citenamefont {{Putman}}, \citenamefont {{Ryder}}, \citenamefont {{Sadler}}, \citenamefont {{Stewart}}, \citenamefont
  {{Stootman}},\ and\ \citenamefont {{Wright}}}]{2004MNRAS.350.1195M}%
  \BibitemOpen
  \bibfield  {author} {\bibinfo {author} {\bibfnamefont {M.~J.}\ \bibnamefont {{Meyer}}}, \bibinfo {author} {\bibfnamefont {M.~A.}\ \bibnamefont {{Zwaan}}}, \bibinfo {author} {\bibfnamefont {R.~L.}\ \bibnamefont {{Webster}}}, \bibinfo {author} {\bibfnamefont {L.}~\bibnamefont {{Staveley-Smith}}}, \bibinfo {author} {\bibfnamefont {E.}~\bibnamefont {{Ryan-Weber}}}, \bibinfo {author} {\bibfnamefont {M.~J.}\ \bibnamefont {{Drinkwater}}}, \bibinfo {author} {\bibfnamefont {D.~G.}\ \bibnamefont {{Barnes}}}, \bibinfo {author} {\bibfnamefont {M.}~\bibnamefont {{Howlett}}}, \emph {et~al.},\ }\href {https://doi.org/10.1111/j.1365-2966.2004.07710.x} {\bibfield  {journal} {\bibinfo  {journal} {\mnras}\ }\textbf {\bibinfo {volume} {350}},\ \bibinfo {pages} {1195} (\bibinfo {year} {2004})},\ \Eprint {https://arxiv.org/abs/astro-ph/0406384} {arXiv:astro-ph/0406384 [astro-ph]} \BibitemShut {NoStop}%
\bibitem [{\citenamefont {{Koss}}\ \emph {et~al.}(2022a)\citenamefont {{Koss}}, \citenamefont {{Trakhtenbrot}}, \citenamefont {{Ricci}}, \citenamefont {{Oh}}, \citenamefont {{Bauer}}, \citenamefont {{Stern}}, \citenamefont {{Caglar}}, \citenamefont {{den Brok}}, \citenamefont {{Mushotzky}}, \citenamefont {{Ricci}}, \citenamefont {{Mej{\'\i}a-Restrepo}}, \citenamefont {{Lamperti}}, \citenamefont {{Treister}}, \citenamefont {{B{\"a}r}}, \citenamefont {{Harrison}}, \citenamefont {{Powell}}, \citenamefont {{Privon}}, \citenamefont {{Riffel}}, \citenamefont {{Rojas}}, \citenamefont {{Schawinski}},\ and\ \citenamefont {{Urry}}}]{2022ApJS..261....6K}%
  \BibitemOpen
  \bibfield  {author} {\bibinfo {author} {\bibfnamefont {M.~J.}\ \bibnamefont {{Koss}}}, \bibinfo {author} {\bibfnamefont {B.}~\bibnamefont {{Trakhtenbrot}}}, \bibinfo {author} {\bibfnamefont {C.}~\bibnamefont {{Ricci}}}, \bibinfo {author} {\bibfnamefont {K.}~\bibnamefont {{Oh}}}, \bibinfo {author} {\bibfnamefont {F.~E.}\ \bibnamefont {{Bauer}}}, \bibinfo {author} {\bibfnamefont {D.}~\bibnamefont {{Stern}}}, \bibinfo {author} {\bibfnamefont {T.}~\bibnamefont {{Caglar}}}, \bibinfo {author} {\bibfnamefont {J.~S.}\ \bibnamefont {{den Brok}}}, \emph {et~al.},\ }\href {https://doi.org/10.3847/1538-4365/ac650b} {\bibfield  {journal} {\bibinfo  {journal} {\apjs}\ }\textbf {\bibinfo {volume} {261}},\ \bibinfo {eid} {6} (\bibinfo {year} {2022})},\ \Eprint {https://arxiv.org/abs/2207.12435} {arXiv:2207.12435 [astro-ph.GA]} \BibitemShut {NoStop}%
\bibitem [{\citenamefont {{Mori}}\ \emph {et~al.}(2015)\citenamefont {{Mori}}, \citenamefont {{Hailey}}, \citenamefont {{Krivonos}}, \citenamefont {{Hong}}, \citenamefont {{Ponti}}, \citenamefont {{Bauer}}, \citenamefont {{Perez}}, \citenamefont {{Nynka}}, \citenamefont {{Zhang}}, \citenamefont {{Tomsick}}, \citenamefont {{Alexander}}, \citenamefont {{Baganoff}}, \citenamefont {{Barret}}, \citenamefont {{Barri{\`e}re}}, \citenamefont {{Boggs}}, \citenamefont {{Canipe}}, \citenamefont {{Christensen}}, \citenamefont {{Craig}}, \citenamefont {{Forster}}, \citenamefont {{Giommi}}, \citenamefont {{Grefenstette}}, \citenamefont {{Grindlay}}, \citenamefont {{Harrison}}, \citenamefont {{Hornstrup}}, \citenamefont {{Kitaguchi}}, \citenamefont {{Koglin}}, \citenamefont {{Luu}}, \citenamefont {{Madsen}}, \citenamefont {{Mao}}, \citenamefont {{Miyasaka}}, \citenamefont {{Perri}}, \citenamefont {{Pivovaroff}}, \citenamefont {{Puccetti}}, \citenamefont {{Rana}}, \citenamefont {{Stern}}, \citenamefont {{Westergaard}},
  \citenamefont {{Zhang}},\ and\ \citenamefont {{Zoglauer}}}]{mytorus_tbabs}%
  \BibitemOpen
  \bibfield  {author} {\bibinfo {author} {\bibfnamefont {K.}~\bibnamefont {{Mori}}}, \bibinfo {author} {\bibfnamefont {C.~J.}\ \bibnamefont {{Hailey}}}, \bibinfo {author} {\bibfnamefont {R.}~\bibnamefont {{Krivonos}}}, \bibinfo {author} {\bibfnamefont {J.}~\bibnamefont {{Hong}}}, \bibinfo {author} {\bibfnamefont {G.}~\bibnamefont {{Ponti}}}, \bibinfo {author} {\bibfnamefont {F.}~\bibnamefont {{Bauer}}}, \bibinfo {author} {\bibfnamefont {K.}~\bibnamefont {{Perez}}}, \bibinfo {author} {\bibfnamefont {M.}~\bibnamefont {{Nynka}}}, \emph {et~al.},\  }\href {https://doi.org/10.1088/0004-637X/814/2/94} {\bibfield  {journal} {\bibinfo  {journal} {\apj}\ }\textbf {\bibinfo {volume} {814}},\ \bibinfo {eid} {94} (\bibinfo {year} {2015})},\ \Eprint {https://arxiv.org/abs/1510.04631} {arXiv:1510.04631 [astro-ph.HE]} \BibitemShut {NoStop}%
\bibitem [{\citenamefont {{Yaqoob}}(2012)}]{mytorus}%
  \BibitemOpen
  \bibfield  {author} {\bibinfo {author} {\bibfnamefont {T.}~\bibnamefont {{Yaqoob}}},\ }\href {https://doi.org/10.1111/j.1365-2966.2012.21129.x} {\bibfield  {journal} {\bibinfo  {journal} {\mnras}\ }\textbf {\bibinfo {volume} {423}},\ \bibinfo {pages} {3360} (\bibinfo {year} {2012})},\ \Eprint {https://arxiv.org/abs/1204.4196} {arXiv:1204.4196 [astro-ph.HE]} \BibitemShut {NoStop}%
\bibitem [{\citenamefont {{Paiano}}\ \emph {et~al.}(2018)\citenamefont {{Paiano}}, \citenamefont {{Falomo}}, \citenamefont {{Treves}},\ and\ \citenamefont {{Scarpa}}}]{2018ApJ...854L..32P}%
  \BibitemOpen
  \bibfield  {author} {\bibinfo {author} {\bibfnamefont {S.}~\bibnamefont {{Paiano}}}, \bibinfo {author} {\bibfnamefont {R.}~\bibnamefont {{Falomo}}}, \bibinfo {author} {\bibfnamefont {A.}~\bibnamefont {{Treves}}},\ and\ \bibinfo {author} {\bibfnamefont {R.}~\bibnamefont {{Scarpa}}},\ }\href {https://doi.org/10.3847/2041-8213/aaad5e} {\bibfield  {journal} {\bibinfo  {journal} {\apjl}\ }\textbf {\bibinfo {volume} {854}},\ \bibinfo {eid} {L32} (\bibinfo {year} {2018})},\ \Eprint {https://arxiv.org/abs/1802.01939} {arXiv:1802.01939 [astro-ph.GA]} \BibitemShut {NoStop}%
\bibitem [{\citenamefont {{Dickey}}\ and\ \citenamefont {{Lockman}}(1990)}]{1990ARA&A..28..215D}%
  \BibitemOpen
  \bibfield  {author} {\bibinfo {author} {\bibfnamefont {J.~M.}\ \bibnamefont {{Dickey}}}\ and\ \bibinfo {author} {\bibfnamefont {F.~J.}\ \bibnamefont {{Lockman}}},\ }\href {https://doi.org/10.1146/annurev.aa.28.090190.001243} {\bibfield  {journal} {\bibinfo  {journal} {\araa}\ }\textbf {\bibinfo {volume} {28}},\ \bibinfo {pages} {215} (\bibinfo {year} {1990})}\BibitemShut {NoStop}%
\bibitem [{\citenamefont {{Kalberla}}\ \emph {et~al.}(2005)\citenamefont {{Kalberla}}, \citenamefont {{Burton}}, \citenamefont {{Hartmann}}, \citenamefont {{Arnal}}, \citenamefont {{Bajaja}}, \citenamefont {{Morras}},\ and\ \citenamefont {{P{\"o}ppel}}}]{2005A&A...440..775K}%
  \BibitemOpen
  \bibfield  {author} {\bibinfo {author} {\bibfnamefont {P.~M.~W.}\ \bibnamefont {{Kalberla}}}, \bibinfo {author} {\bibfnamefont {W.~B.}\ \bibnamefont {{Burton}}}, \bibinfo {author} {\bibfnamefont {D.}~\bibnamefont {{Hartmann}}}, \bibinfo {author} {\bibfnamefont {E.~M.}\ \bibnamefont {{Arnal}}}, \bibinfo {author} {\bibfnamefont {E.}~\bibnamefont {{Bajaja}}}, \bibinfo {author} {\bibfnamefont {R.}~\bibnamefont {{Morras}}},\ and\ \bibinfo {author} {\bibfnamefont {W.~G.~L.}\ \bibnamefont {{P{\"o}ppel}}},\ }\href {https://doi.org/10.1051/0004-6361:20041864} {\bibfield  {journal} {\bibinfo  {journal} {\aap}\ }\textbf {\bibinfo {volume} {440}},\ \bibinfo {pages} {775} (\bibinfo {year} {2005})},\ \Eprint {https://arxiv.org/abs/astro-ph/0504140} {arXiv:astro-ph/0504140 [astro-ph]} \BibitemShut {NoStop}%
\bibitem [{\citenamefont {{HI4PI Collaboration}}\ \emph {et~al.}(2016)\citenamefont {{HI4PI Collaboration}}, \citenamefont {{Ben Bekhti}}, \citenamefont {{Fl{\"o}er}}, \citenamefont {{Keller}}, \citenamefont {{Kerp}}, \citenamefont {{Lenz}}, \citenamefont {{Winkel}}, \citenamefont {{Bailin}}, \citenamefont {{Calabretta}}, \citenamefont {{Dedes}}, \citenamefont {{Ford}}, \citenamefont {{Gibson}}, \citenamefont {{Haud}}, \citenamefont {{Janowiecki}}, \citenamefont {{Kalberla}}, \citenamefont {{Lockman}}, \citenamefont {{McClure-Griffiths}}, \citenamefont {{Murphy}}, \citenamefont {{Nakanishi}}, \citenamefont {{Pisano}},\ and\ \citenamefont {{Staveley-Smith}}}]{2016A&A...594A.116H}%
  \BibitemOpen
  \bibfield  {author} {\bibinfo {author} {\bibnamefont {{HI4PI Collaboration}}}, \bibinfo {author} {\bibfnamefont {N.}~\bibnamefont {{Ben Bekhti}}}, \bibinfo {author} {\bibfnamefont {L.}~\bibnamefont {{Fl{\"o}er}}}, \bibinfo {author} {\bibfnamefont {R.}~\bibnamefont {{Keller}}}, \bibinfo {author} {\bibfnamefont {J.}~\bibnamefont {{Kerp}}}, \bibinfo {author} {\bibfnamefont {D.}~\bibnamefont {{Lenz}}}, \bibinfo {author} {\bibfnamefont {B.}~\bibnamefont {{Winkel}}}, \bibinfo {author} {\bibfnamefont {J.}~\bibnamefont {{Bailin}}}, \bibinfo {author} {\bibfnamefont {M.~R.}\ \bibnamefont {{Calabretta}}}, \emph {et~al.},\ }\href {https://doi.org/10.1051/0004-6361/201629178} {\bibfield  {journal} {\bibinfo  {journal} {\aap}\ }\textbf {\bibinfo {volume} {594}},\ \bibinfo {eid} {A116} (\bibinfo {year} {2016})},\ \Eprint {https://arxiv.org/abs/1610.06175} {arXiv:1610.06175 [astro-ph.GA]} \BibitemShut {NoStop}%
\bibitem [{\citenamefont {{Harrison}}\ \emph {et~al.}(2013)\citenamefont {{Harrison}}, \citenamefont {{Craig}}, \citenamefont {{Christensen}}, \citenamefont {{Hailey}}, \citenamefont {{Zhang}}, \citenamefont {{Boggs}}, \citenamefont {{Stern}}, \citenamefont {{Cook}}, \citenamefont {{Forster}}, \citenamefont {{Giommi}}, \citenamefont {{Grefenstette}}, \citenamefont {{Kim}}, \citenamefont {{Kitaguchi}}, \citenamefont {{Koglin}}, \citenamefont {{Madsen}}, \citenamefont {{Mao}}, \citenamefont {{Miyasaka}}, \citenamefont {{Mori}}, \citenamefont {{Perri}}, \citenamefont {{Pivovaroff}}, \citenamefont {{Puccetti}}, \citenamefont {{Rana}}, \citenamefont {{Westergaard}}, \citenamefont {{Willis}}, \citenamefont {{Zoglauer}}, \citenamefont {{An}}, \citenamefont {{Bachetti}}, \citenamefont {{Barri{\`e}re}}, \citenamefont {{Bellm}}, \citenamefont {{Bhalerao}}, \citenamefont {{Brejnholt}}, \citenamefont {{Fuerst}}, \citenamefont {{Liebe}}, \citenamefont {{Markwardt}}, \citenamefont {{Nynka}}, \citenamefont {{Vogel}},
  \citenamefont {{Walton}}, \citenamefont {{Wik}}, \citenamefont {{Alexander}}, \citenamefont {{Cominsky}}, \citenamefont {{Hornschemeier}}, \citenamefont {{Hornstrup}}, \citenamefont {{Kaspi}}, \citenamefont {{Madejski}}, \citenamefont {{Matt}}, \citenamefont {{Molendi}}, \citenamefont {{Smith}}, \citenamefont {{Tomsick}}, \citenamefont {{Ajello}}, \citenamefont {{Ballantyne}}, \citenamefont {{Balokovi{\'c}}}, \citenamefont {{Barret}}, \citenamefont {{Bauer}}, \citenamefont {{Blandford}}, \citenamefont {{Brandt}}, \citenamefont {{Brenneman}}, \citenamefont {{Chiang}}, \citenamefont {{Chakrabarty}}, \citenamefont {{Chenevez}}, \citenamefont {{Comastri}}, \citenamefont {{Dufour}}, \citenamefont {{Elvis}}, \citenamefont {{Fabian}}, \citenamefont {{Farrah}}, \citenamefont {{Fryer}}, \citenamefont {{Gotthelf}}, \citenamefont {{Grindlay}}, \citenamefont {{Helfand}}, \citenamefont {{Krivonos}}, \citenamefont {{Meier}}, \citenamefont {{Miller}}, \citenamefont {{Natalucci}}, \citenamefont {{Ogle}}, \citenamefont
  {{Ofek}}, \citenamefont {{Ptak}}, \citenamefont {{Reynolds}}, \citenamefont {{Rigby}}, \citenamefont {{Tagliaferri}}, \citenamefont {{Thorsett}}, \citenamefont {{Treister}},\ and\ \citenamefont {{Urry}}}]{2013ApJ...770..103H}%
  \BibitemOpen
  \bibfield  {author} {\bibinfo {author} {\bibfnamefont {F.~A.}\ \bibnamefont {{Harrison}}}, \bibinfo {author} {\bibfnamefont {W.~W.}\ \bibnamefont {{Craig}}}, \bibinfo {author} {\bibfnamefont {F.~E.}\ \bibnamefont {{Christensen}}}, \bibinfo {author} {\bibfnamefont {C.~J.}\ \bibnamefont {{Hailey}}}, \bibinfo {author} {\bibfnamefont {W.~W.}\ \bibnamefont {{Zhang}}}, \bibinfo {author} {\bibfnamefont {S.~E.}\ \bibnamefont {{Boggs}}}, \bibinfo {author} {\bibfnamefont {D.}~\bibnamefont {{Stern}}}, \bibinfo {author} {\bibfnamefont {W.~R.}\ \bibnamefont {{Cook}}}, \emph {et~al.},\ }\href {https://doi.org/10.1088/0004-637X/770/2/103} {\bibfield  {journal} {\bibinfo  {journal} {\apj}\ }\textbf {\bibinfo {volume} {770}},\ \bibinfo {eid} {103} (\bibinfo {year} {2013})},\ \Eprint {https://arxiv.org/abs/1301.7307} {arXiv:1301.7307 [astro-ph.IM]} \BibitemShut {NoStop}%
\bibitem [{\citenamefont {{Bauer}}\ \emph {et~al.}(2015)\citenamefont {{Bauer}}, \citenamefont {{Ar{\'e}valo}}, \citenamefont {{Walton}}, \citenamefont {{Koss}}, \citenamefont {{Puccetti}}, \citenamefont {{Gandhi}}, \citenamefont {{Stern}}, \citenamefont {{Alexander}}, \citenamefont {{Balokovi{\'c}}}, \citenamefont {{Boggs}}, \citenamefont {{Brandt}}, \citenamefont {{Brightman}}, \citenamefont {{Christensen}}, \citenamefont {{Comastri}}, \citenamefont {{Craig}}, \citenamefont {{Del Moro}}, \citenamefont {{Hailey}}, \citenamefont {{Harrison}}, \citenamefont {{Hickox}}, \citenamefont {{Luo}}, \citenamefont {{Markwardt}}, \citenamefont {{Marinucci}}, \citenamefont {{Matt}}, \citenamefont {{Rigby}}, \citenamefont {{Rivers}}, \citenamefont {{Saez}}, \citenamefont {{Treister}}, \citenamefont {{Urry}},\ and\ \citenamefont {{Zhang}}}]{2015ApJ...812..116B}%
  \BibitemOpen
  \bibfield  {author} {\bibinfo {author} {\bibfnamefont {F.~E.}\ \bibnamefont {{Bauer}}}, \bibinfo {author} {\bibfnamefont {P.}~\bibnamefont {{Ar{\'e}valo}}}, \bibinfo {author} {\bibfnamefont {D.~J.}\ \bibnamefont {{Walton}}}, \bibinfo {author} {\bibfnamefont {M.~J.}\ \bibnamefont {{Koss}}}, \bibinfo {author} {\bibfnamefont {S.}~\bibnamefont {{Puccetti}}}, \bibinfo {author} {\bibfnamefont {P.}~\bibnamefont {{Gandhi}}}, \bibinfo {author} {\bibfnamefont {D.}~\bibnamefont {{Stern}}}, \bibinfo {author} {\bibfnamefont {D.~M.}\ \bibnamefont {{Alexander}}}, \emph {et~al.},\ }\href {https://doi.org/10.1088/0004-637X/812/2/116} {\bibfield  {journal}
  {\bibinfo  {journal} {\apj}\ }\textbf {\bibinfo {volume} {812}},\ \bibinfo {eid} {116} (\bibinfo {year} {2015})},\ \Eprint {https://arxiv.org/abs/1411.0670} {arXiv:1411.0670 [astro-ph.HE]} \BibitemShut {NoStop}%
\bibitem [{\citenamefont {{Marchesi}}\ \emph {et~al.}(2018)\citenamefont {{Marchesi}}, \citenamefont {{Ajello}}, \citenamefont {{Marcotulli}}, \citenamefont {{Comastri}}, \citenamefont {{Lanzuisi}},\ and\ \citenamefont {{Vignali}}}]{Marchesi2018}%
  \BibitemOpen
  \bibfield  {author} {\bibinfo {author} {\bibfnamefont {S.}~\bibnamefont {{Marchesi}}}, \bibinfo {author} {\bibfnamefont {M.}~\bibnamefont {{Ajello}}}, \bibinfo {author} {\bibfnamefont {L.}~\bibnamefont {{Marcotulli}}}, \bibinfo {author} {\bibfnamefont {A.}~\bibnamefont {{Comastri}}}, \bibinfo {author} {\bibfnamefont {G.}~\bibnamefont {{Lanzuisi}}},\ and\ \bibinfo {author} {\bibfnamefont {C.}~\bibnamefont {{Vignali}}},\ }\href {https://doi.org/10.3847/1538-4357/aaa410} {\bibfield  {journal} {\bibinfo  {journal} {\apj}\ }\textbf {\bibinfo {volume} {854}},\ \bibinfo {eid} {49} (\bibinfo {year} {2018})},\ \Eprint {https://arxiv.org/abs/1801.03166} {arXiv:1801.03166 [astro-ph.HE]} \BibitemShut {NoStop}%
\bibitem [{\citenamefont {{Georgantopoulos}}\ and\ \citenamefont {{Akylas}}(2019)}]{2019A&A...621A..28G}%
  \BibitemOpen
  \bibfield  {author} {\bibinfo {author} {\bibfnamefont {I.}~\bibnamefont {{Georgantopoulos}}}\ and\ \bibinfo {author} {\bibfnamefont {A.}~\bibnamefont {{Akylas}}},\ }\href {https://doi.org/10.1051/0004-6361/201833038} {\bibfield  {journal} {\bibinfo  {journal} {\aap}\ }\textbf {\bibinfo {volume} {621}},\ \bibinfo {eid} {A28} (\bibinfo {year} {2019})},\ \Eprint {https://arxiv.org/abs/1809.03747} {arXiv:1809.03747 [astro-ph.GA]} \BibitemShut {NoStop}%
\bibitem [{\citenamefont {{Pedlar}}\ \emph {et~al.}(1992)\citenamefont {{Pedlar}}, \citenamefont {{Howley}}, \citenamefont {{Axon}},\ and\ \citenamefont {{Unger}}}]{1992MNRAS.259..369P}%
  \BibitemOpen
  \bibfield  {author} {\bibinfo {author} {\bibfnamefont {A.}~\bibnamefont {{Pedlar}}}, \bibinfo {author} {\bibfnamefont {P.}~\bibnamefont {{Howley}}}, \bibinfo {author} {\bibfnamefont {D.~J.}\ \bibnamefont {{Axon}}},\ and\ \bibinfo {author} {\bibfnamefont {S.~W.}\ \bibnamefont {{Unger}}},\ }\href {https://doi.org/10.1093/mnras/259.2.369} {\bibfield  {journal} {\bibinfo  {journal} {\mnras}\ }\textbf {\bibinfo {volume} {259}},\ \bibinfo {pages} {369} (\bibinfo {year} {1992})}\BibitemShut {NoStop}%
\bibitem [{\citenamefont {{Zoghbi}}\ \emph {et~al.}(2019)\citenamefont {{Zoghbi}}, \citenamefont {{Miller}},\ and\ \citenamefont {{Cackett}}}]{2019ApJ...884...26Z}%
  \BibitemOpen
  \bibfield  {author} {\bibinfo {author} {\bibfnamefont {A.}~\bibnamefont {{Zoghbi}}}, \bibinfo {author} {\bibfnamefont {J.~M.}\ \bibnamefont {{Miller}}},\ and\ \bibinfo {author} {\bibfnamefont {E.}~\bibnamefont {{Cackett}}},\ }\href {https://doi.org/10.3847/1538-4357/ab3e31} {\bibfield  {journal} {\bibinfo  {journal} {\apj}\ }\textbf {\bibinfo {volume} {884}},\ \bibinfo {eid} {26} (\bibinfo {year} {2019})},\ \Eprint {https://arxiv.org/abs/1908.09862} {arXiv:1908.09862 [astro-ph.HE]} \BibitemShut {NoStop}%
\bibitem [{\citenamefont {{Gianolli}}\ \emph {et~al.}(2023)\citenamefont {{Gianolli}}, \citenamefont {{Kim}}, \citenamefont {{Bianchi}}, \citenamefont {{Ag{\'\i}s-Gonz{\'a}lez}}, \citenamefont {{Madejski}}, \citenamefont {{Marin}}, \citenamefont {{Marinucci}}, \citenamefont {{Matt}}, \citenamefont {{Middei}}, \citenamefont {{Petrucci}}, \citenamefont {{Soffitta}}, \citenamefont {{Tagliacozzo}}, \citenamefont {{Tombesi}}, \citenamefont {{Ursini}}, \citenamefont {{Barnouin}}, \citenamefont {{De Rosa}}, \citenamefont {{Di Gesu}}, \citenamefont {{Ingram}}, \citenamefont {{Loktev}}, \citenamefont {{Panagiotou}}, \citenamefont {{Podgorny}}, \citenamefont {{Poutanen}}, \citenamefont {{Puccetti}}, \citenamefont {{Ratheesh}}, \citenamefont {{Veledina}}, \citenamefont {{Zhang}}, \citenamefont {{Agudo}}, \citenamefont {{Antonelli}}, \citenamefont {{Bachetti}}, \citenamefont {{Baldini}}, \citenamefont {{Baumgartner}}, \citenamefont {{Bellazzini}}, \citenamefont {{Bongiorno}}, \citenamefont {{Bonino}}, \citenamefont
  {{Brez}}, \citenamefont {{Bucciantini}}, \citenamefont {{Capitanio}}, \citenamefont {{Castellano}}, \citenamefont {{Cavazzuti}}, \citenamefont {{Chen}}, \citenamefont {{Ciprini}}, \citenamefont {{Costa}}, \citenamefont {{Del Monte}}, \citenamefont {{Di Lalla}}, \citenamefont {{Di Marco}}, \citenamefont {{Donnarumma}}, \citenamefont {{Doroshenko}}, \citenamefont {{Dov{\v{c}}iak}}, \citenamefont {{Ehlert}}, \citenamefont {{Enoto}}, \citenamefont {{Evangelista}}, \citenamefont {{Fabiani}}, \citenamefont {{Ferrazzoli}}, \citenamefont {{Garc{\'\i}a}}, \citenamefont {{Gunji}}, \citenamefont {{Heyl}}, \citenamefont {{Iwakiri}}, \citenamefont {{Jorstad}}, \citenamefont {{Kaaret}}, \citenamefont {{Karas}}, \citenamefont {{Kislat}}, \citenamefont {{Kitaguchi}}, \citenamefont {{Kolodziejczak}}, \citenamefont {{Krawczynski}}, \citenamefont {{La Monaca}}, \citenamefont {{Latronico}}, \citenamefont {{Liodakis}}, \citenamefont {{Maldera}}, \citenamefont {{Manfreda}}, \citenamefont {{Marscher}}, \citenamefont {{Marshall}},
  \citenamefont {{Massaro}}, \citenamefont {{Mitsuishi}}, \citenamefont {{Mizuno}}, \citenamefont {{Muleri}}, \citenamefont {{Negro}}, \citenamefont {{Ng}}, \citenamefont {{O'Dell}}, \citenamefont {{Omodei}}, \citenamefont {{Oppedisano}}, \citenamefont {{Papitto}}, \citenamefont {{Pavlov}}, \citenamefont {{Peirson}}, \citenamefont {{Perri}}, \citenamefont {{Pesce-Rollins}}, \citenamefont {{Pilia}}, \citenamefont {{Possenti}}, \citenamefont {{Ramsey}}, \citenamefont {{Rankin}}, \citenamefont {{Roberts}}, \citenamefont {{Romani}}, \citenamefont {{Sgr{\`o}}}, \citenamefont {{Slane}}, \citenamefont {{Spandre}}, \citenamefont {{Swartz}}, \citenamefont {{Tamagawa}}, \citenamefont {{Tavecchio}}, \citenamefont {{Taverna}}, \citenamefont {{Tawara}}, \citenamefont {{Tennant}}, \citenamefont {{Thomas}}, \citenamefont {{Trois}}, \citenamefont {{Tsygankov}}, \citenamefont {{Turolla}}, \citenamefont {{Vink}}, \citenamefont {{Weisskopf}}, \citenamefont {{Wu}}, \citenamefont {{Xie}},\ and\ \citenamefont
  {{Zane}}}]{2023MNRAS.523.4468G}%
  \BibitemOpen
  \bibfield  {author} {\bibinfo {author} {\bibfnamefont {V.~E.}\ \bibnamefont {{Gianolli}}}, \bibinfo {author} {\bibfnamefont {D.~E.}\ \bibnamefont {{Kim}}}, \bibinfo {author} {\bibfnamefont {S.}~\bibnamefont {{Bianchi}}}, \bibinfo {author} {\bibfnamefont {B.}~\bibnamefont {{Ag{\'\i}s-Gonz{\'a}lez}}}, \bibinfo {author} {\bibfnamefont {G.}~\bibnamefont {{Madejski}}}, \bibinfo {author} {\bibfnamefont {F.}~\bibnamefont {{Marin}}}, \bibinfo {author} {\bibfnamefont {A.}~\bibnamefont {{Marinucci}}}, \bibinfo {author} {\bibfnamefont {G.}~\bibnamefont {{Matt}}}, \emph {et~al.},\ }\href {https://doi.org/10.1093/mnras/stad1697} {\bibfield  {journal} {\bibinfo  {journal} {\mnras}\ }\textbf {\bibinfo {volume} {523}},\ \bibinfo {pages} {4468} (\bibinfo {year} {2023})},\ \Eprint {https://arxiv.org/abs/2303.12541} {arXiv:2303.12541 [astro-ph.GA]} \BibitemShut {NoStop}%
\bibitem [{\citenamefont {{Koss}}\ \emph {et~al.}(2022b)\citenamefont {{Koss}}, \citenamefont {{Trakhtenbrot}}, \citenamefont {{Ricci}}, \citenamefont {{Oh}}, \citenamefont {{Bauer}}, \citenamefont {{Stern}}, \citenamefont {{Caglar}}, \citenamefont {{den Brok}}, \citenamefont {{Mushotzky}}, \citenamefont {{Ricci}}, \citenamefont {{Mej{\'\i}a-Restrepo}}, \citenamefont {{Lamperti}}, \citenamefont {{Treister}}, \citenamefont {{B{\"a}r}}, \citenamefont {{Harrison}}, \citenamefont {{Powell}}, \citenamefont {{Privon}}, \citenamefont {{Riffel}}, \citenamefont {{Rojas}}, \citenamefont {{Schawinski}},\ and\ \citenamefont {{Urry}}}]{2022ApJS..261....2K}%
  \BibitemOpen
  \bibfield  {author} {\bibinfo {author} {\bibfnamefont {M.~J.}\ \bibnamefont {{Koss}}}, \bibinfo {author} {\bibfnamefont {B.}~\bibnamefont {{Trakhtenbrot}}}, \bibinfo {author} {\bibfnamefont {C.}~\bibnamefont {{Ricci}}}, \bibinfo {author} {\bibfnamefont {K.}~\bibnamefont {{Oh}}}, \bibinfo {author} {\bibfnamefont {F.~E.}\ \bibnamefont {{Bauer}}}, \bibinfo {author} {\bibfnamefont {D.}~\bibnamefont {{Stern}}}, \bibinfo {author} {\bibfnamefont {T.}~\bibnamefont {{Caglar}}}, \bibinfo {author} {\bibfnamefont {J.~S.}\ \bibnamefont {{den Brok}}}, \emph {et~al.},\ }\href {https://doi.org/ 
10.3847/1538-4365/ac6c05} {\bibfield  {journal} {\bibinfo  {journal} {\apjs}\ }\textbf {\bibinfo {volume} {261}},\ \bibinfo {eid} {2} (\bibinfo {year} {2022})},\ \Eprint {https://arxiv.org/abs/2207.12432} {arXiv:2207.12432 [astro-ph.GA]} \BibitemShut {NoStop}%
\bibitem [{\citenamefont {{Arnaud}}(1996)}]{1996ASPC..101...17A}%
  \BibitemOpen
  \bibfield  {author} {\bibinfo {author} {\bibfnamefont {K.~A.}\ \bibnamefont {{Arnaud}}},\ }in\ \href@noop {} {\emph {\bibinfo {booktitle} {Astronomical Data Analysis Software and Systems V}}},\ \bibinfo {series} {Astronomical Society of the Pacific Conference Series}, Vol.\ \bibinfo {volume} {101},\ \bibinfo {editor} {edited by\ \bibinfo {editor} {\bibfnamefont {G.~H.}\ \bibnamefont {{Jacoby}}}\ and\ \bibinfo {editor} {\bibfnamefont {J.}~\bibnamefont {{Barnes}}}}\ (\bibinfo {year} {1996})\ p.~\bibinfo {pages} {17}\BibitemShut {NoStop}%
\bibitem [{\citenamefont {{Wilms}}\ \emph {et~al.}(2000)\citenamefont {{Wilms}}, \citenamefont {{Allen}},\ and\ \citenamefont {{McCray}}}]{wilm}%
  \BibitemOpen
  \bibfield  {author} {\bibinfo {author} {\bibfnamefont {J.}~\bibnamefont {{Wilms}}}, \bibinfo {author} {\bibfnamefont {A.}~\bibnamefont {{Allen}}},\ and\ \bibinfo {author} {\bibfnamefont {R.}~\bibnamefont {{McCray}}},\ }\href {https://doi.org/10.1086/317016} {\bibfield  {journal} {\bibinfo  {journal} {\apj}\ }\textbf {\bibinfo {volume} {542}},\ \bibinfo {pages} {914} (\bibinfo {year} {2000})},\ \Eprint {https://arxiv.org/abs/astro-ph/0008425} {arXiv:astro-ph/0008425 [astro-ph]} \BibitemShut {NoStop}%
\bibitem [{\citenamefont {{Acciari}}\ \emph {et~al.}(2022)\citenamefont {{Acciari}}, \citenamefont {{Aniello}}, \citenamefont {{Ansoldi}}, \citenamefont {{Antonelli}}, \citenamefont {{Arbet Engels}}, \citenamefont {{Artero}}, \citenamefont {{Asano}}, \citenamefont {{Baack}}, \citenamefont {{Babi{\'c}}}, \citenamefont {{Baquero}}, \citenamefont {{Barres de Almeida}}, \citenamefont {{Barrio}}, \citenamefont {{Batkovi{\'c}}}, \citenamefont {{Becerra Gonz{\'a}lez}}, \citenamefont {{Bednarek}}, \citenamefont {{Bernardini}}, \citenamefont {{Bernardos}}, \citenamefont {{Berti}}, \citenamefont {{Besenrieder}}, \citenamefont {{Bhattacharyya}}, \citenamefont {{Bigongiari}}, \citenamefont {{Biland}}, \citenamefont {{Blanch}}, \citenamefont {{B{\"o}kenkamp}}, \citenamefont {{Bonnoli}}, \citenamefont {{Bo{\v{s}}njak}}, \citenamefont {{Busetto}}, \citenamefont {{Carosi}}, \citenamefont {{Ceribella}}, \citenamefont {{Cerruti}}, \citenamefont {{Chai}}, \citenamefont {{Chilingarian}}, \citenamefont {{Cikota}}, \citenamefont
  {{Colombo}}, \citenamefont {{Contreras}}, \citenamefont {{Cortina}}, \citenamefont {{Covino}}, \citenamefont {{D'Amico}}, \citenamefont {{D'Elia}}, \citenamefont {{Vela}}, \citenamefont {{Dazzi}}, \citenamefont {{De Angelis}}, \citenamefont {{De Lotto}}, \citenamefont {{Del Popolo}}, \citenamefont {{Delfino}}, \citenamefont {{Delgado}}, \citenamefont {{Mendez}}, \citenamefont {{Depaoli}}, \citenamefont {{Di Pierro}}, \citenamefont {{Di Venere}}, \citenamefont {{Do Souto Espi{\~n}eira}}, \citenamefont {{Dominis Prester}}, \citenamefont {{Donini}}, \citenamefont {{Dorner}}, \citenamefont {{Doro}}, \citenamefont {{Elsaesser}}, \citenamefont {{Fallah Ramazani}}, \citenamefont {{Fari{\~n}a}}, \citenamefont {{Fattorini}}, \citenamefont {{Font}}, \citenamefont {{Fruck}}, \citenamefont {{Fukami}}, \citenamefont {{Fukazawa}}, \citenamefont {{Garc{\'\i}a L{\'o}pez}}, \citenamefont {{Garczarczyk}}, \citenamefont {{Gasparyan}}, \citenamefont {{Gaug}}, \citenamefont {{Giglietto}}, \citenamefont {{Giordano}},
  \citenamefont {{Gliwny}}, \citenamefont {{Godinovi{\'c}}}, \citenamefont {{Green}}, \citenamefont {{Green}}, \citenamefont {{Hadasch}}, \citenamefont {{Hahn}}, \citenamefont {{Hassan}}, \citenamefont {{Heckmann}}, \citenamefont {{Herrera}}, \citenamefont {{Hoang}}, \citenamefont {{Hrupec}}, \citenamefont {{H{\"u}tten}}, \citenamefont {{Inada}}, \citenamefont {{Iotov}}, \citenamefont {{Ishio}}, \citenamefont {{Iwamura}}, \citenamefont {{Jim{\'e}nez Mart{\'\i}nez}}, \citenamefont {{Jormanainen}}, \citenamefont {{Jouvin}}, \citenamefont {{Kerszberg}}, \citenamefont {{Kobayashi}}, \citenamefont {{Kubo}}, \citenamefont {{Kushida}}, \citenamefont {{Lamastra}}, \citenamefont {{Lelas}}, \citenamefont {{Leone}}, \citenamefont {{Lindfors}}, \citenamefont {{Linhoff}}, \citenamefont {{Lombardi}}, \citenamefont {{Longo}}, \citenamefont {{L{\'o}pez-Coto}}, \citenamefont {{L{\'o}pez-Moya}}, \citenamefont {{L{\'o}pez-Oramas}}, \citenamefont {{Loporchio}}, \citenamefont {{Machado de Oliveira Fraga}}, \citenamefont
  {{Maggio}}, \citenamefont {{Majumdar}}, \citenamefont {{Makariev}}, \citenamefont {{Mallamaci}}, \citenamefont {{Maneva}}, \citenamefont {{Manganaro}}, \citenamefont {{Mannheim}}, \citenamefont {{Mariotti}}, \citenamefont {{Mart{\'\i}nez}}, \citenamefont {{Mas Aguilar}}, \citenamefont {{Mazin}}, \citenamefont {{Menchiari}}, \citenamefont {{Mender}}, \citenamefont {{Mi{\'c}anovi{\'c}}}, \citenamefont {{Miceli}}, \citenamefont {{Miener}}, \citenamefont {{Miranda}}, \citenamefont {{Mirzoyan}}, \citenamefont {{Molina}}, \citenamefont {{Moralejo}}, \citenamefont {{Morcuende}}, \citenamefont {{Moreno}}, \citenamefont {{Moretti}}, \citenamefont {{Nakamori}}, \citenamefont {{Nava}}, \citenamefont {{Neustroev}}, \citenamefont {{Nievas Rosillo}}, \citenamefont {{Nigro}}, \citenamefont {{Nilsson}}, \citenamefont {{Nishijima}}, \citenamefont {{Noda}}, \citenamefont {{Nozaki}}, \citenamefont {{Ohtani}}, \citenamefont {{Oka}}, \citenamefont {{Otero-Santos}}, \citenamefont {{Paiano}}, \citenamefont {{Palatiello}},
  \citenamefont {{Paneque}}, \citenamefont {{Paoletti}}, \citenamefont {{Paredes}}, \citenamefont {{Pavleti{\'c}}}, \citenamefont {{Pe{\~n}il}}, \citenamefont {{Persic}}, \citenamefont {{Pihet}}, \citenamefont {{Prada Moroni}}, \citenamefont {{Prandini}}, \citenamefont {{Priyadarshi}}, \citenamefont {{Puljak}}, \citenamefont {{Rhode}}, \citenamefont {{Rib{\'o}}}, \citenamefont {{Rico}}, \citenamefont {{Righi}}, \citenamefont {{Rugliancich}}, \citenamefont {{Sahakyan}}, \citenamefont {{Saito}}, \citenamefont {{Sakurai}}, \citenamefont {{Satalecka}}, \citenamefont {{Saturni}}, \citenamefont {{Schleicher}}, \citenamefont {{Schmidt}}, \citenamefont {{Schmuckermaier}}, \citenamefont {{Schweizer}}, \citenamefont {{Sitarek}}, \citenamefont {{{\v{S}}nidari{\'c}}}, \citenamefont {{Sobczynska}}, \citenamefont {{Spolon}}, \citenamefont {{Stamerra}}, \citenamefont {{Stri{\v{s}}kovi{\'c}}}, \citenamefont {{Strom}}, \citenamefont {{Strzys}}, \citenamefont {{Suda}}, \citenamefont {{Suri{\'c}}}, \citenamefont {{Takahashi}},
  \citenamefont {{Takeishi}}, \citenamefont {{Tavecchio}}, \citenamefont {{Temnikov}}, \citenamefont {{Terzi{\'c}}}, \citenamefont {{Teshima}}, \citenamefont {{Tosti}}, \citenamefont {{Truzzi}}, \citenamefont {{Tutone}}, \citenamefont {{Ubach}}, \citenamefont {{van Scherpenberg}}, \citenamefont {{Vanzo}}, \citenamefont {{Vazquez Acosta}}, \citenamefont {{Ventura}}, \citenamefont {{Verguilov}}, \citenamefont {{Viale}}, \citenamefont {{Vigorito}}, \citenamefont {{Vitale}}, \citenamefont {{Vovk}}, \citenamefont {{Will}}, \citenamefont {{Wunderlich}}, \citenamefont {{Yamamoto}}, \citenamefont {{Zari{\'c}}}, \citenamefont {{Hodges}}, \citenamefont {{Hovatta}}, \citenamefont {{Kiehlmann}}, \citenamefont {{Liodakis}}, \citenamefont {{Max-Moerbeck}}, \citenamefont {{Pearson}}, \citenamefont {{Readhead}}, \citenamefont {{Reeves}}, \citenamefont {{L{\"a}hteenm{\"a}ki}}, \citenamefont {{Tornikoski}}, \citenamefont {{Tammi}}, \citenamefont {{D'Ammando}},\ and\ \citenamefont {{Marchini}}}]{variability}%
  \BibitemOpen
  \bibfield  {author} {\bibinfo {author} {\bibfnamefont {V.~A.}\ \bibnamefont {{Acciari}}}, \bibinfo {author} {\bibfnamefont {T.}~\bibnamefont {{Aniello}}}, \bibinfo {author} {\bibfnamefont {S.}~\bibnamefont {{Ansoldi}}}, \bibinfo {author} {\bibfnamefont {L.~A.}\ \bibnamefont {{Antonelli}}}, \bibinfo {author} {\bibfnamefont {A.}~\bibnamefont {{Arbet Engels}}}, \bibinfo {author} {\bibfnamefont {M.}~\bibnamefont {{Artero}}}, \bibinfo {author} {\bibfnamefont {K.}~\bibnamefont {{Asano}}}, \bibinfo {author} {\bibfnamefont {D.}~\bibnamefont {{Baack}}}, \emph {et~al.},\ }\href {https://doi.org/10.3847/1538-4357/ac531d} {\bibfield
  {journal} {\bibinfo  {journal} {\apj}\ }\textbf {\bibinfo {volume} {927}},\ \bibinfo {eid} {197} (\bibinfo {year} {2022})},\ \Eprint {https://arxiv.org/abs/2202.02600} {arXiv:2202.02600 [astro-ph.HE]} \BibitemShut {NoStop}%
\bibitem [{\citenamefont {{March{\~a}}}\ and\ \citenamefont {{Caccianiga}}(2013)}]{Marcha2013}%
  \BibitemOpen
  \bibfield  {author} {\bibinfo {author} {\bibfnamefont {M.~J.~M.}~\bibnamefont {{March{\~a}}}}\ and\ \bibinfo {author} {\bibfnamefont {A.}~\bibnamefont {{Caccianiga}}},\ }\href {https://doi.org/10.1093/mnras/stt065} {\bibfield  {journal} {\bibinfo  {journal} {\mnras}\ }\textbf {\bibinfo {volume} {430}},\ \bibinfo {eid} {2464} (\bibinfo {year} {2013})},\ \Eprint {https://arxiv.org/abs/1301.6550} {arXiv:1301.6550 [astro-ph.HE]} \BibitemShut {NoStop}%
\bibitem [{\citenamefont {{Evans}}\ \emph {et~al.}(2007)\citenamefont {{Evans}} \emph {et~al.}}]{Evans2007}%
  \BibitemOpen
  \bibfield  {author} {\bibinfo {author} {\bibfnamefont {P.~A.}\ \bibnamefont {{Evans}}} \emph {et~al.},\ }\href {https://doi.org/10.1051/0004-6361:20077530} {\bibfield  {journal} {\bibinfo  {journal} {\aap}\ }\textbf {\bibinfo {volume} {469}},\ \bibinfo {eid} {379} (\bibinfo {year} {2007})},\ \Eprint {https://arxiv.org/abs/0704.0128} {arXiv:0704.0128 [astro-ph.HE]} \BibitemShut {NoStop}%
\bibitem [{\citenamefont {{Abbasi}}\ \emph {et~al.}(2024{\natexlab{a}})\citenamefont {{Abbasi}}, \citenamefont {{Ackermann}}, \citenamefont {{Adams}}, \citenamefont {{Aguilar}}, \citenamefont {{Ahlers}}, \citenamefont {{Ahrens}}, \citenamefont {{Alameddine}}, \citenamefont {{Alispach}}  \emph {et~al.}}]{Abassi2019}%
  \BibitemOpen
  \bibfield  {author} {\bibinfo {author} {\bibfnamefont {R.}~\bibnamefont {{Abbasi}}}, \bibinfo {author} {\bibfnamefont {M.}~\bibnamefont {{Ackermann}}}, \bibinfo {author} {\bibfnamefont {J.}~\bibnamefont {{Adams}}}, \bibinfo {author} {\bibfnamefont {J.~A.}\ \bibnamefont {{Aguilar}}}, \bibinfo {author} {\bibfnamefont {M.}~\bibnamefont {{Ahlers}}}, \bibinfo {author} {\bibfnamefont {M.}~\bibnamefont {{Ahrens}}}, \bibinfo {author} {\bibfnamefont {J.~M.}\ \bibnamefont {{Alameddine}}}, \bibinfo {author} {\bibfnamefont {C.}~\bibnamefont {{Alispach}}}, \emph {et~al.},\ }\href {https://doi.org/10.48550/arXiv.2101.09836} {\bibfield  {journal} {\bibinfo  {journal} {arXiv e-prints}\ ,\ \bibinfo {eid} {arXiv:2101.09836}} (\bibinfo {year} {2019})},\ \Eprint {https://arxiv.org/abs/2101.09836} {arXiv:2101.09836 [astro-ph.HE]} \BibitemShut {NoStop}%
\bibitem [{\citenamefont {{Ansoldi}}\ \emph {et~al.}(2018)}]{Ansoldi2018}%
  \BibitemOpen
  \bibfield  {author} {\bibinfo {author} {\bibfnamefont {S.}~\bibnamefont {{Ansoldi}}}, \bibinfo {author} {\bibfnamefont {L.~A.}\ \bibnamefont {{Antonelli}}}, \bibinfo {author} {\bibfnamefont {C.}~\bibnamefont {{Arcaro}}}, \bibinfo {author} {\bibfnamefont {D.}~\bibnamefont {{Baack}}}, \bibinfo {author} {\bibfnamefont {A.}~\bibnamefont {{Babi{\'c}}}}, \bibinfo {author} {\bibfnamefont {B.}~\bibnamefont {{Banerjee}}}, \bibinfo {author} {\bibfnamefont {P.}~\bibnamefont {{Bangale}}}, \bibinfo {author} {\bibfnamefont {U.}~\bibnamefont {{Barres de Almeida}}}, \emph {et~al.},\ }\href {https://doi.org/10.3847/2041-8213/aad083} {\bibfield  {journal} {\bibinfo  {journal} {\apjl}\ }\textbf {\bibinfo
  {volume} {863}},\ \bibinfo {eid} {L10} (\bibinfo {year} {2018})},\ \Eprint {https://arxiv.org/abs/1807.04300} {arXiv:1807.04300 [astro-ph.HE]} \BibitemShut {NoStop}%
\bibitem [{\citenamefont {{Halzen}}(2022)}]{2022IJMPD..3130003H}%
  \BibitemOpen
  \bibfield  {author} {\bibinfo {author} {\bibfnamefont {F.}~\bibnamefont {{Halzen}}},\ }\href {https://doi.org/10.1142/S0218271822300038} {\bibfield  {journal} {\bibinfo  {journal} {International Journal of Modern Physics D}\ }\textbf {\bibinfo {volume} {31}},\ \bibinfo {eid} {2230003-8} (\bibinfo {year} {2022})},\ \Eprint {https://arxiv.org/abs/2110.01687} {arXiv:2110.01687 [astro-ph.HE]} \BibitemShut {NoStop}%
  \bibitem [{\citenamefont {{Svensson}}(1987)}]{1987MNRAS.227..403S}%
  \BibitemOpen
  \bibfield  {author} {\bibinfo {author} {\bibfnamefont {R.}~\bibnamefont {{Svensson}}},\ }\href {https://doi.org/10.1093/mnras/227.2.403} {\bibfield  {journal} {\bibinfo  {journal} {\mnras}\ }\textbf {\bibinfo {volume} {227}},\ \bibinfo {pages} {403} (\bibinfo {year} {1987})}\BibitemShut {NoStop}%
\bibitem [{\citenamefont {{Murase}}(2022)}]{2022ApJ...941L..17M}%
  \BibitemOpen
  \bibfield  {author} {\bibinfo {author} {\bibfnamefont {K.}~\bibnamefont {{Murase}}},\ }\href {https://doi.org/10.3847/2041-8213/aca53c} {\bibfield  {journal} {\bibinfo  {journal} {\apjl}\ }\textbf {\bibinfo {volume} {941}},\ \bibinfo {eid} {L17} (\bibinfo {year} {2022})},\ \Eprint {https://arxiv.org/abs/2211.04460} {arXiv:2211.04460 [astro-ph.HE]} \BibitemShut {NoStop}%
\bibitem [{\citenamefont {Gao}\ \emph {et~al.}(2019)\citenamefont {Gao}, \citenamefont {Fedynitch}, \citenamefont {Winter},\ and\ \citenamefont {Pohl}}]{Walter1}%
  \BibitemOpen
  \bibfield  {author} {\bibinfo {author} {\bibfnamefont {S.}~\bibnamefont {Gao}}, \bibinfo {author} {\bibfnamefont {A.}~\bibnamefont {Fedynitch}}, \bibinfo {author} {\bibfnamefont {W.}~\bibnamefont {Winter}},\ and\ \bibinfo {author} {\bibfnamefont {M.}~\bibnamefont {Pohl}},\ }\href {https://doi.org/10.1038/s41550-018-0610-1} {\bibfield  {journal} {\bibinfo  {journal} {Nature Astronomy}\ }\textbf {\bibinfo {volume} {3}},\ \bibinfo {pages} {88} (\bibinfo {year} {2019})}\BibitemShut {NoStop}%
\bibitem [{\citenamefont {{Kun}}\ \emph {et~al.}(2022)\citenamefont {{Kun}}, \citenamefont {{Bartos}}, \citenamefont {{Becker Tjus}}, \citenamefont {{Biermann}}, \citenamefont {{Franckowiak}},\ and\ \citenamefont {{Halzen}}}]{Kun2022MW}%
  \BibitemOpen
  \bibfield  {author} {\bibinfo {author} {\bibfnamefont {E.}~\bibnamefont {{Kun}}}, \bibinfo {author} {\bibfnamefont {I.}~\bibnamefont {{Bartos}}}, \bibinfo {author} {\bibfnamefont {J.}~\bibnamefont {{Becker Tjus}}}, \bibinfo {author} {\bibfnamefont {P.~L.}\ \bibnamefont {{Biermann}}}, \bibinfo {author} {\bibfnamefont {A.}~\bibnamefont {{Franckowiak}}},\ and\ \bibinfo {author} {\bibfnamefont {F.}~\bibnamefont {{Halzen}}},\ }\href {https://doi.org/10.3847/1538-4357/ac7f3a} {\bibfield  {journal} {\bibinfo  {journal} {\apj}\ }\textbf {\bibinfo {volume} {934}},\ \bibinfo {eid} {180} (\bibinfo {year} {2022})},\ \Eprint {https://arxiv.org/abs/2203.14780} {arXiv:2203.14780 [astro-ph.HE]} \BibitemShut {NoStop}%
\bibitem [{\citenamefont {{Aartsen}}\ \emph {et~al.}(2018)\citenamefont {{Aartsen}}, \citenamefont {{Ackermann}}, \citenamefont {{Adams}}, \citenamefont {{Aguilar}}, \citenamefont {{Ahlers}}, \citenamefont {{Ahrens}}, \citenamefont {{Samarai}}, \citenamefont {{Altmann}}, \citenamefont {{Andeen}}, \citenamefont {{Anderson}}, \citenamefont {{Ansseau}}, \citenamefont {{Anton}}, \citenamefont {{Arg{\"u}elles}}, \citenamefont {{Arsioli}}, \citenamefont {{Auffenberg}} \emph {et~al.}}]{2018Sci...361..147I}%
  \BibitemOpen
  \bibfield  {author} {\bibinfo {author} {\bibfnamefont {M.~G.}\ \bibnamefont {{Aartsen}}}, \bibinfo {author} {\bibfnamefont {M.}~\bibnamefont {{Ackermann}}}, \bibinfo {author} {\bibfnamefont {J.}~\bibnamefont {{Adams}}}, \bibinfo {author} {\bibfnamefont {J.~A.}\ \bibnamefont {{Aguilar}}}, \bibinfo {author} {\bibfnamefont {M.}~\bibnamefont {{Ahlers}}}, \bibinfo {author} {\bibfnamefont {M.}~\bibnamefont {{Ahrens}}}, \bibinfo {author} {\bibfnamefont {I.~A.}\ \bibnamefont {{Samarai}}}, \bibinfo {author} {\bibfnamefont {D.}~\bibnamefont {{Altmann}}}, \bibinfo {author} {\bibfnamefont {K.}~\bibnamefont {{Andeen}}}, \bibinfo {author} {\bibfnamefont {T.}~\bibnamefont {{Anderson}}}, \bibinfo {author} {\bibfnamefont {I.}~\bibnamefont {{Ansseau}}}, \bibinfo {author} {\bibfnamefont {G.}~\bibnamefont {{Anton}}}, \bibinfo {author} {\bibfnamefont {C.}~\bibnamefont {{Arg{\"u}elles}}}, \bibinfo {author} {\bibfnamefont {B.}~\bibnamefont {{Arsioli}}}, \bibinfo {author} {\bibfnamefont {J.}~\bibnamefont {{Auffenberg}}}, \emph
  {et~al.},\ }\href {https://doi.org/10.1126/science.aat2890} {\bibfield  {journal} {\bibinfo  {journal} {Science}\ }\textbf {\bibinfo {volume} {361}},\ \bibinfo {pages} {147} (\bibinfo {year} {2018})},\ \Eprint {https://arxiv.org/abs/1807.08794} {arXiv:1807.08794 [astro-ph.HE]} \BibitemShut {NoStop}%
\bibitem [{\citenamefont {{Stecker}}\ and\ \citenamefont {{Salamon}}(1996)}]{1996tgra.conf..341S}%
  \BibitemOpen
  \bibfield  {author} {\bibinfo {author} {\bibfnamefont {F.~W.}\ \bibnamefont {{Stecker}}}\ and\ \bibinfo {author} {\bibfnamefont {M.~H.}\ \bibnamefont {{Salamon}}},\ }in\ \href@noop {} {\emph {\bibinfo {booktitle} {TeV Gamma-ray Astrophysics. Theory and Observations}}},\ \bibinfo {editor} {edited by\ \bibinfo {editor} {\bibfnamefont {H.~J.}\ \bibnamefont {{V{\"o}lk}}}\ and\ \bibinfo {editor} {\bibfnamefont {F.~A.}\ \bibnamefont {{Aharonian}}}}\ (\bibinfo {year} {1996})\ pp.\ \bibinfo {pages} {341--355}\BibitemShut {NoStop}%
\bibitem [{\citenamefont {{Keck}}\ \emph {et~al.}(2015)\citenamefont {{Keck}},
  \citenamefont {{Brenneman}}, \citenamefont {{Ballantyne}}, \citenamefont
  {{Bauer}}, \citenamefont {{Boggs}}, \citenamefont {{Christensen}},
  \citenamefont {{Craig}}, \citenamefont {{Dauser}}, \citenamefont {{Elvis}},
  \citenamefont {{Fabian}}, \citenamefont {{Fuerst}}, \citenamefont
  {{Garc{\'\i}a}}, \citenamefont {{Grefenstette}}, \citenamefont {{Hailey}},
  \citenamefont {{Harrison}}, \citenamefont {{Madejski}}, \citenamefont
  {{Marinucci}}, \citenamefont {{Matt}}, \citenamefont {{Reynolds}},
  \citenamefont {{Stern}}, \citenamefont {{Walton}},\ and\ \citenamefont
  {{Zoghbi}}}]{2015ApJ...806..149K}%
  \BibitemOpen
  \bibfield  {author} {\bibinfo {author} {\bibfnamefont {M.~L.}\ \bibnamefont
  {{Keck}}}, \bibinfo {author} {\bibfnamefont {L.~W.}\ \bibnamefont
  {{Brenneman}}}, \bibinfo {author} {\bibfnamefont {D.~R.}\ \bibnamefont
  {{Ballantyne}}}, \bibinfo {author} {\bibfnamefont {F.}~\bibnamefont
  {{Bauer}}}, \bibinfo {author} {\bibfnamefont {S.~E.}\ \bibnamefont
  {{Boggs}}}, \bibinfo {author} {\bibfnamefont {F.~E.}\ \bibnamefont
  {{Christensen}}}, \bibinfo {author} {\bibfnamefont {W.~W.}\ \bibnamefont
  {{Craig}}}, \bibinfo {author} {\bibfnamefont {T.}~\bibnamefont {{Dauser}}},
  \bibinfo {author} {\bibfnamefont {M.}~\bibnamefont {{Elvis}}}, \bibinfo
  {author} {\bibfnamefont {A.~C.}\ \bibnamefont {{Fabian}}}, \bibinfo {author}
  {\bibfnamefont {F.}~\bibnamefont {{Fuerst}}}, \bibinfo {author}
  {\bibfnamefont {J.}~\bibnamefont {{Garc{\'\i}a}}}, \bibinfo {author}
  {\bibfnamefont {B.~W.}\ \bibnamefont {{Grefenstette}}}, \bibinfo {author}
  {\bibfnamefont {C.~J.}\ \bibnamefont {{Hailey}}}, \bibinfo {author}
  {\bibfnamefont {F.~A.}\ \bibnamefont {{Harrison}}}, \bibinfo {author}
  {\bibfnamefont {G.}~\bibnamefont {{Madejski}}}, \bibinfo {author}
  {\bibfnamefont {A.}~\bibnamefont {{Marinucci}}}, \bibinfo {author}
  {\bibfnamefont {G.}~\bibnamefont {{Matt}}}, \bibinfo {author} {\bibfnamefont
  {C.~S.}\ \bibnamefont {{Reynolds}}}, \bibinfo {author} {\bibfnamefont
  {D.}~\bibnamefont {{Stern}}}, \bibinfo {author} {\bibfnamefont {D.~J.}\
  \bibnamefont {{Walton}}},\ and\ \bibinfo {author} {\bibfnamefont
  {A.}~\bibnamefont {{Zoghbi}}},\ }\bibfield  {title} {\bibinfo {title}
  {{NuSTAR and Suzaku X-ray Spectroscopy of NGC 4151: Evidence for Reflection
  from the Inner Accretion Disk}},\ }\href
  {https://doi.org/10.1088/0004-637X/806/2/149} {\bibfield  {journal} {\bibinfo
   {journal} {\apj}\ }\textbf {\bibinfo {volume} {806}},\ \bibinfo {eid} {149}
  (\bibinfo {year} {2015})},\ \Eprint {https://arxiv.org/abs/1504.07950}
  {arXiv:1504.07950 [astro-ph.HE]} \BibitemShut {NoStop}%
\bibitem [{\citenamefont {{Gopal-Krishna}}\ and\ \citenamefont {{Biermann}}(2024)}]{2024MNRAS.529L.135G}%
  \BibitemOpen
  \bibfield  {author} {\bibinfo {author} {\bibnamefont {{Gopal-Krishna}}}\ and\ \bibinfo {author} {\bibfnamefont {P.~L.}\ \bibnamefont {{Biermann}}},\ }\href {https://doi.org/10.1093/mnrasl/slad191} {\bibfield  {journal} {\bibinfo  {journal} {\mnras}\ }\textbf {\bibinfo {volume} {529}},\ \bibinfo {pages} {L135} (\bibinfo {year} {2024})}\BibitemShut {NoStop}%
\bibitem [{\citenamefont {{Kardashev}}(1962)}]{1962SvA.....6..317K}%
  \BibitemOpen
  \bibfield  {author} {\bibinfo {author} {\bibfnamefont {N.~S.}\ \bibnamefont {{Kardashev}}},\ }\href@noop {} {\bibfield  {journal} {\bibinfo  {journal} {\sovast}\ }\textbf {\bibinfo {volume} {6}},\ \bibinfo {pages} {317} (\bibinfo {year} {1962})}\BibitemShut {NoStop}%
\bibitem [{\citenamefont {{Abdollahi}}\ \emph {et~al.}(2022a)\citenamefont {{Abdollahi}}, \citenamefont {{Ajello}}, \citenamefont {{Baldini}}, \citenamefont {{Ballet}}, \citenamefont {{Bastieri}}, \citenamefont {{Becerra Gonzalez}}, \citenamefont {{Bellazzini}}, \citenamefont {{Berretta}}, \citenamefont {{Mushotzky}}, \citenamefont {{Ricci}}, \citenamefont {{Mej{\'\i}a-Restrepo}}, \citenamefont {{Lamperti}}, \citenamefont {{Treister}}, \citenamefont {{B{\"a}r}}, \citenamefont {{Harrison}}, \citenamefont {{Powell}}, \citenamefont {{Privon}}, \citenamefont {{Riffel}}, \citenamefont {{Rojas}}, \citenamefont {{Schawinski}},\ and\ \citenamefont {{Urry}}}]{Abdollahi2023}%
  \BibitemOpen
  \bibfield  {author} {\bibinfo {author} {\bibfnamefont {S.}\ \bibnamefont {{Abdollahi}}}, \bibinfo {author} {\bibfnamefont {M.}~\bibnamefont {{Ajello}}}, \bibinfo {author} {\bibfnamefont {L.}~\bibnamefont {{Baldini}}}, \bibinfo {author} {\bibfnamefont {J.}~\bibnamefont {{Ballet}}}, \bibinfo {author} {\bibfnamefont {D.}\ \bibnamefont {{Bastieri}}}, \bibinfo {author} {\bibfnamefont {J.}~\bibnamefont {{Becerra Gonzalez}}}, \bibinfo {author} {\bibfnamefont {R.}~\bibnamefont {{Bellazzini}}}, \bibinfo {author} {\bibfnamefont {A.}\ \bibnamefont {{Berretta}}}, \emph {et~al.},\ }\href {https://doi.org/10.3847/1538-4365/acbb6a} {\bibfield  {journal} {\bibinfo  {journal} {\apjs}\ }\textbf {\bibinfo {volume} {265}},\ \bibinfo {eid} {31} (\bibinfo {year} {2023})} \BibitemShut {NoStop}%
\bibitem [{\citenamefont {{Astropy Collaboration}}\ \emph {et~overoveal.}(2013)\citenamefont {{Astropy Collaboration}}, \citenamefont {{Robitaille}}, \citenamefont {{Tollerud}}, \citenamefont {{Greenfield}}, \citenamefont {{Droettboom}}, \citenamefont {{Bray}}, \citenamefont {{Aldcroft}}, \citenamefont {{Davis}}, \citenamefont {{Ginsburg}}, \citenamefont {{Price-Whelan}}, \citenamefont {{Kerzendorf}}, \citenamefont {{Conley}}, \citenamefont {{Crighton}}, \citenamefont {{Barbary}}, \citenamefont {{Muna}}, \citenamefont {{Ferguson}}, \citenamefont {{Grollier}}, \citenamefont {{Parikh}}, \citenamefont {{Nair}}, \citenamefont {{Unther}}, \citenamefont {{Deil}}, \citenamefont {{Woillez}}, \citenamefont {{Conseil}}, \citenamefont {{Kramer}}, \citenamefont {{Turner}}, \citenamefont {{Singer}}, \citenamefont {{Fox}}, \citenamefont {{Weaver}}, \citenamefont {{Zabalza}}, \citenamefont {{Edwards}}, \citenamefont {{Azalee Bostroem}}, \citenamefont {{Burke}}, \citenamefont {{Casey}}, \citenamefont {{Crawford}}, \citenamefont
  {{Dencheva}}, \citenamefont {{Ely}}, \citenamefont {{Jenness}}, \citenamefont {{Labrie}}, \citenamefont {{Lim}}, \citenamefont {{Pierfederici}}, \citenamefont {{Pontzen}}, \citenamefont {{Ptak}}, \citenamefont {{Refsdal}}, \citenamefont {{Servillat}},\ and\ \citenamefont {{Streicher}}}]{astropy:2013}%
  \BibitemOpen
  \bibfield  {author} {\bibinfo {author} {\bibnamefont {{Astropy Collaboration}}}, \bibinfo {author} {\bibfnamefont {T.~P.}\ \bibnamefont {{Robitaille}}}, \bibinfo {author} {\bibfnamefont {E.~J.}\ \bibnamefont {{Tollerud}}}, \bibinfo {author} {\bibfnamefont {P.}~\bibnamefont {{Greenfield}}}, \bibinfo {author} {\bibfnamefont {M.}~\bibnamefont {{Droettboom}}}, \bibinfo {author} {\bibfnamefont {E.}~\bibnamefont {{Bray}}}, \bibinfo {author} {\bibfnamefont {T.}~\bibnamefont {{Aldcroft}}}, \bibinfo {author} {\bibfnamefont {M.}~\bibnamefont {{Davis}}}, \bibinfo {author} {\bibfnamefont {A.}~\bibnamefont {{Ginsburg}}}, \emph {et~al.},\ }\href {https://doi.org/10.1051/0004-6361/201322068}
  {\bibfield  {journal} {\bibinfo  {journal} {\aap}\ }\textbf {\bibinfo {volume} {558}},\ \bibinfo {eid} {A33} (\bibinfo {year} {2013})},\ \Eprint {https://arxiv.org/abs/1307.6212} {arXiv:1307.6212 [astro-ph.IM]} \BibitemShut {NoStop}%
\bibitem [{\citenamefont {{Astropy Collaboration}}\ \emph {et~al.}(2022)\citenamefont {{Astropy Collaboration}}, \citenamefont {{Price-Whelan}}, \citenamefont {{Lim}}, \citenamefont {{Earl}}, \citenamefont {{Starkman}}, \citenamefont {{Bradley}}, \citenamefont {{Shupe}}, \citenamefont {{Patil}}, \citenamefont {{Corrales}}, \citenamefont {{Brasseur}}, \citenamefont {{N{\"o}the}}, \citenamefont {{Donath}}, \citenamefont {{Tollerud}}, \citenamefont {{Morris}}, \citenamefont {{Ginsburg}}, \citenamefont {{Vaher}}, \citenamefont {{Weaver}}, \citenamefont {{Tocknell}}, \citenamefont {{Jamieson}}, \citenamefont {{van Kerkwijk}}, \citenamefont {{Robitaille}}, \citenamefont {{Merry}}, \citenamefont {{Bachetti}}, \citenamefont {{G{\"u}nther}}, \citenamefont {{Aldcroft}}, \citenamefont {{Alvarado-Montes}}, \citenamefont {{Archibald}}, \citenamefont {{B{\'o}di}}, \citenamefont {{Bapat}}, \citenamefont {{Barentsen}}, \citenamefont {{Baz{\'a}n}}, \citenamefont {{Biswas}}, \citenamefont {{Boquien}}, \citenamefont {{Burke}},
  \citenamefont {{Cara}}, \citenamefont {{Cara}}, \citenamefont {{Conroy}}, \citenamefont {{Conseil}}, \citenamefont {{Craig}}, \citenamefont {{Cross}}, \citenamefont {{Cruz}}, \citenamefont {{D'Eugenio}}, \citenamefont {{Dencheva}}, \citenamefont {{Devillepoix}}, \citenamefont {{Dietrich}}, \citenamefont {{Eigenbrot}}, \citenamefont {{Erben}}, \citenamefont {{Ferreira}}, \citenamefont {{Foreman-Mackey}}, \citenamefont {{Fox}}, \citenamefont {{Freij}}, \citenamefont {{Garg}}, \citenamefont {{Geda}}, \citenamefont {{Glattly}}, \citenamefont {{Gondhalekar}}, \citenamefont {{Gordon}}, \citenamefont {{Grant}}, \citenamefont {{Greenfield}}, \citenamefont {{Groener}}, \citenamefont {{Guest}}, \citenamefont {{Gurovich}}, \citenamefont {{Handberg}}, \citenamefont {{Hart}}, \citenamefont {{Hatfield-Dodds}}, \citenamefont {{Homeier}}, \citenamefont {{Hosseinzadeh}}, \citenamefont {{Jenness}}, \citenamefont {{Jones}}, \citenamefont {{Joseph}}, \citenamefont {{Kalmbach}}, \citenamefont {{Karamehmetoglu}}, \citenamefont
  {{Ka{\l}uszy{\'n}ski}}, \citenamefont {{Kelley}}, \citenamefont {{Kern}}, \citenamefont {{Kerzendorf}}, \citenamefont {{Koch}}, \citenamefont {{Kulumani}}, \citenamefont {{Lee}}, \citenamefont {{Ly}}, \citenamefont {{Ma}}, \citenamefont {{MacBride}}, \citenamefont {{Maljaars}}, \citenamefont {{Muna}}, \citenamefont {{Murphy}}, \citenamefont {{Norman}}, \citenamefont {{O'Steen}}, \citenamefont {{Oman}}, \citenamefont {{Pacifici}}, \citenamefont {{Pascual}}, \citenamefont {{Pascual-Granado}}, \citenamefont {{Patil}}, \citenamefont {{Perren}}, \citenamefont {{Pickering}}, \citenamefont {{Rastogi}}, \citenamefont {{Roulston}}, \citenamefont {{Ryan}}, \citenamefont {{Rykoff}}, \citenamefont {{Sabater}}, \citenamefont {{Sakurikar}}, \citenamefont {{Salgado}}, \citenamefont {{Sanghi}}, \citenamefont {{Saunders}}, \citenamefont {{Savchenko}}, \citenamefont {{Schwardt}}, \citenamefont {{Seifert-Eckert}}, \citenamefont {{Shih}}, \citenamefont {{Jain}}, \citenamefont {{Shukla}}, \citenamefont {{Sick}}, \citenamefont
  {{Simpson}}, \citenamefont {{Singanamalla}}, \citenamefont {{Singer}}, \citenamefont {{Singhal}}, \citenamefont {{Sinha}}, \citenamefont {{Sip{\H{o}}cz}}, \citenamefont {{Spitler}}, \citenamefont {{Stansby}}, \citenamefont {{Streicher}}, \citenamefont {{{\v{S}}umak}}, \citenamefont {{Swinbank}}, \citenamefont {{Taranu}}, \citenamefont {{Tewary}}, \citenamefont {{Tremblay}}, \citenamefont {{de Val-Borro}}, \citenamefont {{Van Kooten}}, \citenamefont {{Vasovi{\'c}}}, \citenamefont {{Verma}}, \citenamefont {{de Miranda Cardoso}}, \citenamefont {{Williams}}, \citenamefont {{Wilson}}, \citenamefont {{Winkel}}, \citenamefont {{Wood-Vasey}}, \citenamefont {{Xue}}, \citenamefont {{Yoachim}}, \citenamefont {{Zhang}}, \citenamefont {{Zonca}},\ and\ \citenamefont {{Astropy Project Contributors}}}]{astropy:2022}%
  \BibitemOpen
  
  \bibfield  {author} {\bibinfo {author} {\bibnamefont {{Astropy Collaboration}}}, \bibinfo {author} {\bibfnamefont {A.~M.}\ \bibnamefont {{Price-Whelan}}}, \bibinfo {author} {\bibfnamefont {P.~L.}\ \bibnamefont {{Lim}}}, \bibinfo {author} {\bibfnamefont {N.}~\bibnamefont {{Earl}}}, \bibinfo {author} {\bibfnamefont {N.}~\bibnamefont {{Starkman}}}, \bibinfo {author} {\bibfnamefont {L.}~\bibnamefont {{Bradley}}}, \bibinfo {author} {\bibfnamefont {D.~L.}\ \bibnamefont {{Shupe}}}, \bibinfo {author} {\bibfnamefont {A.~A.}\ \bibnamefont {{Patil}}}, \bibinfo {author} {\bibfnamefont {L.}~\bibnamefont {{Corrales}}}, \emph {et~al.},\ }\href {https://doi.org/10.3847/1538-4357/ac7c74} {\bibfield  {journal} {\bibinfo  {journal} {\apj}\ }\textbf {\bibinfo {volume} {935}},\ \bibinfo {eid} {167} (\bibinfo {year} {2022})},\ \Eprint {https://arxiv.org/abs/2206.14220} {arXiv:2206.14220 [astro-ph.IM]} \BibitemShut {NoStop}%
\end{thebibliography}
\end{document}